\documentclass[journal,english]{IEEEtran}

\usepackage{comment}

\usepackage{epsfig}
\usepackage{times}
\usepackage{float}
\usepackage{afterpage}

\usepackage{amsmath}
\usepackage{amstext,cite}
\usepackage{amssymb,bm}
\usepackage{latexsym}
\usepackage{color}
\usepackage{graphicx}
\usepackage{amsmath}
\usepackage{amsthm}
\usepackage{graphicx}
\usepackage[center]{caption}
\usepackage{pstricks}
\usepackage{caption}
\usepackage{subcaption}
\usepackage{booktabs}
\usepackage{multicol}
\usepackage{lipsum}
\usepackage[T1]{fontenc}
\usepackage{hyperref}
\usepackage{aecompl}
\usepackage{bbm}

\usepackage{empheq}
 \usepackage{mathrsfs}
\allowdisplaybreaks

\newtheorem{theorem}{Theorem}
\newtheorem{remark}{Remark}

\long\def\comment#1{}

\newcommand{\CC}{{\mathbb C}}

\newcommand{\hv}{{\mathbf h}}

\newcommand{\vv}{{\mathbf v}}
\newcommand{\xv}{{\mathbf x}}
\newcommand{\yv}{{\mathbf y}}
\newcommand{\zv}{{\mathbf z}}

\newcommand{\Am}{{\mathbf A}}

\newcommand{\Hm}{{\mathbf H}}

\newcommand{\Mm}{{\mathbf M}}

\newcommand{\Sm}{{\mathbf S}}

\newcommand{\Wm}{{\mathbf W}}

\newcommand{\Xm}{{\mathbf X}}
\newcommand{\Ym}{{\mathbf Y}}

\newcommand{\Ac}{{\mathcal A}}

\newcommand{\Rc}{{\mathcal R}}

\newcommand{\asf}{{\mathsf a}}
\newcommand{\bsf}{{\mathsf b}}

\newcommand{\msf}{{\mathsf m}}
\newcommand{\nsf}{{\mathsf n}}

\newcommand{\psf}{{\mathsf p}}
\newcommand{\qsf}{{\mathsf q}}
\newcommand{\rsf}{{\mathsf r}}

\newcommand{\Ksf}{{\mathsf K}}

\newcommand{\Ssf}{{\mathsf S}}

\renewcommand{\arg}{{\hbox{arg}}}

\newtheorem{example}{Example}
\newtheorem{defn}{\protect\definitionname}

\providecommand{\definitionname}{Definition}

\def\BibTeX{{\rm B\kern-.05em{\sc i\kern-.025em b}\kern-.08em
    T\kern-.1667em\lower.7ex\hbox{E}\kern-.125emX}}

\begin{document}

\title{Blind and Topological Interference Managements for Bistatic Integrated Sensing and Communication}

\author{
Jiayu~Liu, 
Kai~Wan,~\IEEEmembership{Member,~IEEE,} 
Xinping~Yi,~\IEEEmembership{Member,~IEEE,}
Robert~Caiming~Qiu,~\IEEEmembership{Fellow,~IEEE},
and~Giuseppe Caire,~\IEEEmembership{Fellow,~IEEE}
\thanks{
A short version of this paper   was accepted by  the 2025 IEEE International Conference on Communications. 
}
\thanks{J.~Liu, K.~Wan, and R.~C.~Qiu are  with the School of Electronic Information and Communications, 
Huazhong University of Science and Technology, 430074  Wuhan, China  (e-mail: \{jiayuliu,kai\_wan,caiming\}@hust.edu.cn).}
\thanks{
X.~Yi is with the Southeast University, Nanjing 210096, China (e-mail:xyi@seu.edu.cn).}
\thanks{
 G.~Caire is with the Electrical Engineering and Computer
Science Department, Technische Universit\"at Berlin, 10587 Berlin, Germany
(e-mail:  caire@tu-berlin.de).}
}

\maketitle

\begin{abstract}
Integrated sensing and communication (ISAC) systems
provide significant enhancements in
performance and resource efficiency compared to individual sensing and communication systems, primarily attributed to the collaborative use of wireless resources, radio waveforms, and hardware platforms. 

This paper focuses on the bistatic ISAC systems with 
dispersed multi-receiver and one sensor. 
Compared to a monostatic ISAC system, the main challenge in the bistatic setting is that the information messages are unknown to the sensor and therefore they are seen as interference, while the channel between the transmitters (TX) and the sensor is unknown to the transmitters. In order to mitigate the interference at the sensor while 
maximizing the communication degree of freedom, we introduce two strategies, namely,  blind interference alignment and topological interference management. Although well-known in the context of Gaussian interference channels,  these strategies are novel in the context of bistatic ISAC.  

For the bistatic ISAC models with heterogeneous coherence times or with heterogeneous connectivity, 
the achieved ISAC tradeoff points in terms of  communication and sensing degrees of freedom  are characterized. In particular, we show that the new tradeoff outperforms the time-sharing between the sensing-only and the communication-only schemes. Simulation results  demonstrate  that the proposed schemes  significantly improve the channel estimation error for the sensing task, compared to treating interference as noise at the sensor and successive interference cancellation. 
\end{abstract}


\section{Introduction}

Driven by the continuous evolution of wireless communication technologies, future wireless networks are required to support both high-speed communication over wide-area coverage and large-scale sensing capabilities, thereby enabling high-precision modeling and real-time monitoring of the physical environment \cite{liu2022integrated}. With the fast growth of communication terminals and radar devices, the coexistence of communication and sensing systems under limited spectrum resources has become increasingly prominent \cite{liu2018mu}. To address this challenge, Integrated Sensing and Communication (ISAC) has emerged,   achieving dual-system synergy through innovative mechanisms such as hardware sharing and waveform fusion.

\begin{figure}
    \centering
    \includegraphics[width=0.4\textwidth]{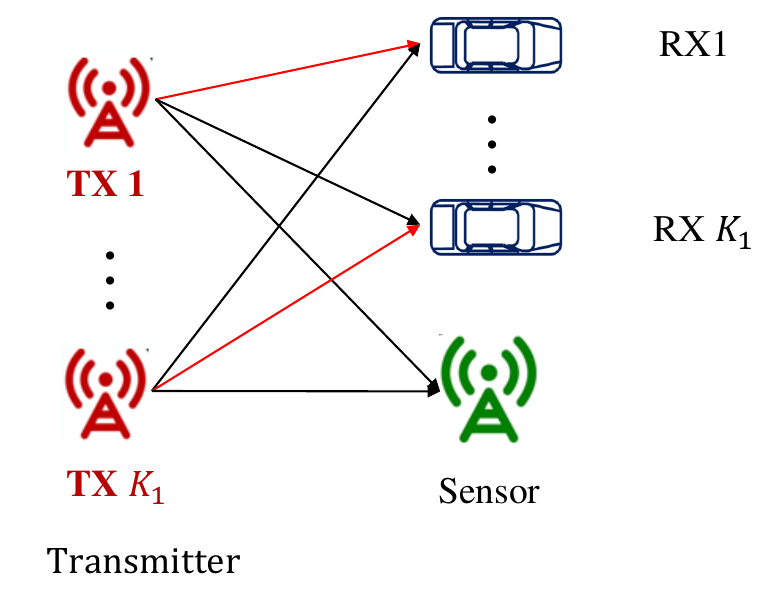}
    \caption{\small The considered bistatic ISAC systems}.
    \label{fig:senior}
\end{figure}

Despite the fundamental differences in information processing principles between sensing and communication, significant research efforts have been devoted to exploring the essence of ``integration'' gains in ISAC systems. These studies focus on the information-theoretic trade-offs between communication and sensing performance, aiming to establish a theoretical foundation for their co-design. The information-theoretic framework for ISAC was first introduced in~\cite{Kobayashi2018ISAC,ahmadipour2022information}, where the authors investigated a monostatic ISAC system with state-dependent discrete memoryless channels and an independent and identically distributed (i.i.d.) state sequence. The capacity-distortion region was characterized by maximizing conditional mutual information under specific constraints. To explicitly derive the closed-form on the trade-off between communication and sensing performances, Xiong {\it et al.}~\cite{xiong2023fundamental} analyzed the Multiple-Input Multiple-Output (MIMO)
 Gaussian channel and explored the Cramér-Rao Bound (CRB)-rate region, which identified two extreme trade-off points: the communication-optimal point and the sensing-optimal point, which further revealed two critical aspects of ISAC trade-offs: subspace trade-offs and deterministic-random trade-offs. For scenarios with fixed sensing states, where the state remains constant over a transmission block, the optimal trade-off between communication rate and state detection error exponent was characterized in~\cite{joudeh2022joint}. Monostatic ISAC over interference channels was considered in~\cite{liu2024infor,liu2022fundamental}, where  two interfering ISAC transmitters 
  communicate with their own users or a common user while  sensing
estimation through received echo signals, and an improved random coding than the superposition coding  for
interference channels was proposed. 
Readers can refer to the surveys in~\cite{liu2022survey,Xiong2024BITS} for more details on the information theoretic progress for ISAC. 

Most existing ISAC research has focused on monostatic systems, where sensing tasks exploit communication signals as side information. In addition, the monostatic architecture inherently suffers from strong self-interference and limited spatial coverage. In contrast, bistatic sensing—--where the sensor and transmitters are spatially separated—can avoid direct reflections, exploit multipath propagation, and enhance detection of low-altitude or ground targets~\cite{willis2007advances}.
From an information-theoretic perspective\footnote{{\label{foot::avoidradar}}The authors are fully aware of the fact that a whole host of problems related to the synchronization (time, frequency, phase) of transmitters and sensors play an important role in bistatic sensing. Nevertheless, these problems can be circumvented by appropriate system design. For example, we may imagine that base stations and sensor are part of the same network running a high precision synchronization protocol, which effectively eliminate the problem to the required degree of accuracy. In contrast, conveying the information bits sent to the users by multiple base stations without a suitable delay is a much harder task, since these messages depend on local (at the base station) scheduling decisions and the throughput of such traffic would be huge. Hence, it is reasonable to assume that the information messages are not know at the sensor, while the synchronization issues have been reduced to a degree that they can be neglected for the sake of analytical tractability. }, research on bistatic ISAC mainly concentrates on two scenarios depending on the relative positions of communication receivers(RX) and sensors. 
\begin{itemize}
    \item For bistatic ISAC models with colocated receiver and sensor, the work~\cite{zhang2011joint} characterizes the capacity-distortion region for i.i.d. state sequences, where the optimal strategy for the sensor is to first decode the communication message and then proceed with the estimation, referred to as  unique decoding. 
The authors in~\cite{chang2023ratedetection} proposed an achievable region for the  fixed state case by  using composition
codes and  joint decoding-detection strategy, since in this case the above unique decoding is sub-optimal. The authors studied the bistatic ISAC model for the joint multi-target localization and data communication task in~\cite{zhao2024joint}, by designing a combination strategy of the pilot and data symbols. The bistatic ISAC model with colocated receiver and sensor over relay channels was considered in~\cite{liu2024bistatic}.  
\item For the bistatic ISAC model with dispersed receivers and sensor,  
 a key challenge arises from the fact that communication messages are not necessary to be decoded by the sensor and thus act as its interference.\footnote{\label{foot:bistatic}In this paper, we mainly focus on how to cancel the interference from the communication message to the sensor. Thus another existing bistatic ISAC model with dispersed receivers and sensor, where the sensor knows the communication message (such as in~\cite{nikbakht2025mimo}), is out of scope.}
 The authors in~\cite{jiao2023information} developed three achievable decoding-and-estimation strategies for sensors,  blind estimation (e.g., by treating  interference  as noise as in~\cite{dong2023joint}),  by partial-decoding-based estimation (e.g., by the successive interference cancellation (SIC) as in~\cite{xu2021rate,ouyang2023revealing}), and full-decoding-based estimation (i.e., fully decoding the communication message and then estimating). 
  Interference mitigation via sharing radar sampling information and joint optimization of the transmit covariance matrices and radar sampling strategy was considered in~\cite{li2016optimum}, where the communication-induced interference was effectively mitigated to enable radar-communication coexistence.
  Mutual interference management through joint spatio-temporal transmit sequence design was studied in~\cite{qian2023enhancement}, where practical constraints such as power, Peak-to-Average Power Ratio (PAR), and spectral leakage were incorporated to ensure spectrum compatibility in MIMO coexistence systems. 
Physical secrecy issue was considered in~\cite{gong2024secrecy}, where the sensor cannot obtain any information about the communication message while estimating the state.
\end{itemize}

\paragraph*{Main Contributions}

This paper considers the bistatic ISAC model with multiple transmitters,  dispersed multi-receivers, and one sensor.
In order to characterize the fundamental tradeoff between communication and sensing performance, we consider ``homogeneous'' metrics, communication degree of freedom (cDoF) and sensing degree  of freedom (sDoF),  representing the average number of effective transmissions and observations in each time slot, respectively. Two new  interference management strategies are first introduced into the bistatic ISAC model   in order to  perfectly eliminate interference to the sensor, instead of recovering all the interference as in~\cite{jiao2023information} or reducing the interference as in~\cite{jiao2023information,li2016optimum,qian2023enhancement}.\footnote{\label{foot:IA for ISAC} The interference alignment strategy was  introduced into ISAC systems in~\cite{cui2018perspective}, by using the original interference alignment scheme in~\cite{cadambe2008ia} with the assumption that the   channels from the communication transmitters  and the (transformed) channels from the radar transmitters to the sensor are all known.   In our paper, we assume that any channel to the sensor is unknown.} 
\begin{itemize}
    \item One uses the blind interference alignment (BIA) strategy originally proposed in~\cite{jafar2012blind}. We assume that 
     each transmitter is connected to all the receivers and  sensor through wireless channels,  and the channel between the transmitters and the sensor is slow-fading and unknown to the transmitters, while the channels between the transmitters and communication receivers are fast-fading and known to the transmitters and receivers.\footnote{\label{foot:slow and fast} The practical motivation is twofold. (i) Communication channels vary rapidly due to user mobility, with a coherence time of approximately 1–10 ms. These channels are periodically estimated via UL/DL reciprocity and uplink pilots, from which the transmitters acquires channel state information (CSI). For simplicity, we assume perfect CSI and ignore pilot overhead. (ii) In contrast, sensing channels vary slowly as sensors and targets are typically stationary over short time spans. However, they are often passive and do not transmit pilots or provide feedback, making their CSI unavailable at the transmitters despite the slow variation. This asymmetric CSI assumption—known for communication but unknown for sensing—has also been adopted in information-theoretic ISAC models such as~\cite{xiong2023fundamental,jiao2023information}.
}  
     We consider three types of wireless channels, including interference channel, multi-user multiple-input single-output (MU-MISO) channel, and multi-user MIMO (MU-MIMO) channel.
    \item  The other strategy is the topological interference management (TIM) originally proposed in~\cite{jafar2013topological}.  We follow the topological interference network model in~\cite{jafar2013topological, maleki2014index, yi2015topological}, assuming   that 
     each transmitter is connected to a subset of  the sensor and receivers through slow-fading wireless channels,  while the transmitters only know the prior information of the network topology, instead of the full CSI. 
      In this model,  the   regular network and the neighboring antidotes network are considered, determining   the structure of the network topology.
     
\end{itemize}

 By using the two interference management strategies, BIA and TIM, we propose new tradeoff points (in closed-form) between cDoF and sDoF, which are strictly better than the time-sharing between the two extreme sensing-optimal and communication-optimal points. 
 The common ingredient of using the above two strategies into ISAC is to leverage the channel heterogeneity among the sensor and receivers to align interference at the sensor without knowing the  CSI of the sensor channel, where the  heterogeneity is on the coherence times for the first model, and on the connection topology for the second model.  
  Finally,  we perform simulations on a practical ISAC system with Differential Quadrature Phase Shift Keying (DQPSK) modulation and additive white Gaussian noise. With channel estimation error as the sensing metric, the proposed method outperforms both treating interference as noise (TIN)-based sensing and SIC, which relies on decoding communication signals prior to sensing channel estimation.

 The remainder of this paper is organized as follows. Section~II introduces the system model of the considered ISAC system and the corresponding performance metrics. 
 Section III  summarizes the main results of this paper. 
 Section IV describes the proposed ISAC schemes based on the BIA strategy, and Section V describes the proposed  ISAC schemes based on the TIM strategy. 
 Section VI provides simulation results to validate the performance of the proposed
 schemes. 
 Finally, Section VII concludes the paper.

\paragraph*{Notation Convention}
We adopt boldface letters to refer to vectors  and matrices. Sets are denoted using calligraphic symbols. Sans-serif symbols denote system parameters.
For an arbitrary-size matrix $\Mm$, $\text{rank}(\Mm)$, $\Mm^*$, $\Mm^T$, and $\Mm^H$ represent its rank, conjugate, transpose, and conjugate transpose, respectively. \(\lceil \cdot \rceil\) represents the ceiling function, which denotes rounding up to the nearest integer. The m-by-m identity matrix is denoted by $\textbf{I}_m$.
  \( \Ac^c \) is the complementary set of \( \Ac \), and \( |\Ac| \) is the cardinality of   \( A \). \( \Am_{ij} \) or \( [\Am]_{ij} \) presents the $ (i,j) $-th entry of the matrix \( \Am \). 
If $a$ is not divisible by $b$, $<a >_b$ denotes the least non-negative residue of $a$ modulo $b$; otherwise, $< a>_b:=b$.

\section{System Model}
\subsection{Channel Models}
\label{sub:channel model}
Consider a  bistatic ISAC system, where \(\Ksf_T\)  transmitters are equipped with \(\msf_i\) antennas where \(i \in \mathcal{T} = \{1, \dots, \Ksf_T\}\), and \(\Ksf_R\) communication receivers (simply called receivers), each receiver $k\in \mathcal{R} = \{1, \dots, \Ksf_R\}$  with \(\nsf_k\) antennas. Without loss of generality, assume that $ \nsf_1 \geq \nsf_2 \geq \cdots \geq \nsf_{\Ksf}$.
The sensor is equipped with a single antenna.
The channel outputs at the $\Ksf_R$  receivers and at the sensor at the \( t \)-th time slot are,
\begin{align*}
&    \yv_c^{[k]}(t) = \negmedspace\negmedspace \sum_{i=1}^{\Ksf_T}\negmedspace  g_{ki} (\Hm^{[ki]}(t) \xv^{[i]}(t) + \zv^{[k]}(t)) \in \mathbb{C}^{\nsf_k \times 1}, \ k \in \mathcal{R}, \\
 &   y_s(t) = \sum_{i=1}^{\Ksf_T} g_s (\Hm^{[(\Ksf_R+1)i]}(t) \xv^{[i]}(t) + z_s(t)),
\end{align*}
where \(\Hm^{[ki]}(t) \in \mathbb{C}^{\nsf_k \times \msf_i}\) represents the channel  from the \(i\)-th transmitter to the \(k\)-th receiver at time slot \(t\), and \(\Hm^{[(\Ksf_R+1)i]}(t) \in \mathbb{C}^{1 \times \msf_i}\) represents the channel  from the \(i\)-th transmitter to the sensor at time slot~\(t\);  
\(\zv^{[k]}(t) \in \mathbb{C}^{\nsf_k \times 1}\) and \(z_s(t) \in \mathbb{C}\) are the additive Gaussian white noise at the receivers and sensor, respectively, with each element i.i.d. as   \( \mathcal{CN}(0, 1)\). \( g_{ki} \in \{0,1\} \) 
and \( g_s  \in \{0,1\} \) indicate the connectivity parameters between the transmitters and the receivers/sensor. 
The network connectivity (i.e., topology) is known to all the transmitters,   receivers, and sensor. 
Assume that each transmission block is composed of  \( t_0 \) channel uses. 
 The signal transmitted by each transmitter $i\in \mathcal{T}$ at time slot $t\in \{1,2,\ldots,t_0\}$  is  the sum of a communication signal and  a dedicated sensing signal:
\begin{equation*}
    \xv^{[i]}(t) = \xv_c^{[i]}(t) + \xv_s^{[i]}(t) \in \mathbb{C}^{\msf_i \times 1}.
\end{equation*}
The transmitted signal \(\xv^{[i]}(t)\) satisfies the power constraint \( \mathbb{E}[|\xv^{[i]}(t)|^2] \leq P \). 
  \(\xv_c^{[i]}(t) \in \mathbb{C}^{\msf_i\times 1}\) represents the communication signal, generated by the product of some precoding matrix and some vector of message symbols,\footnote{\label{foot:Gaussian coding}Message symbols are encoded by using the Gaussian encoding with rate $\log P+ o(\log P)$ (bit per message symbol). 
  Thus each message symbol carries one DoF for large enough $P$.} and 
the dedicated sensing signal \( \xv_s^{[i]}(t) \in \mathbb{C}^{\msf \times 1} \) is fixed and thus known to all the receivers and  sensor.  

Let   the unbolded notations
$y_c^{[k]}(t),$ $H^{[ki]}(t),$ $z^{[k]}(t),$ $x^{[i]}(t),$ $x_c^{[i]}(t),$ $x_s^{[i]}(t)$ represent  $\yv_c^{[k]}(t),$ $ \Hm^{[ki]}(t),$ $\zv^{[k]}(t),$ $ \xv^{[i]}(t),$ $\xv_c^{[i]}(t),$ $\xv_s^{[i]}(t)$ each only containing   one element, respectively.


This paper aims to study interference mitigation based on the  channel pattern among the  receivers and the sensor.
We consider  two types of channel models: (i) heterogeneous coherence times, and (ii) 
heterogeneous connectivity.


\subsubsection{Heterogeneous coherence times}
\label{fading channel}
Consider that the transmitters and receivers/sensors are fully connected. 
The communication channel follows a block fading model with a short coherence time, whereas the sensing channel follows a block fading model with a long coherence time.  The channel state information of the communication channel is perfectly known by the transmitters and receivers.\footnote{\label{foot:regular}Through channel estimation and prediction, that is, the trajectories of the receivers movements are regular and predictable.} 
The difference in coherence time arises because the receivers (such as vehicular terminals) experience significant Doppler shifts due to high mobility, leading to a short channel coherence time (on the order of milliseconds). 
In contrast, sensors (such as static sensors) experience negligible Doppler effects due to their low-speed or stationary nature, resulting in a long channel coherence time (tens of milliseconds or more), which supports continuous signal integration to enhance sensing accuracy. Under this scenario, we assume that the communication channel varies in every time slot  following a Rayleigh fading model, while the sensor channel   remains
approximately constant over the given time period $t_0$. 

\begin{figure}[htbp]
    \centering
    \subfloat[]{%
        \includegraphics[width=0.16\textwidth]{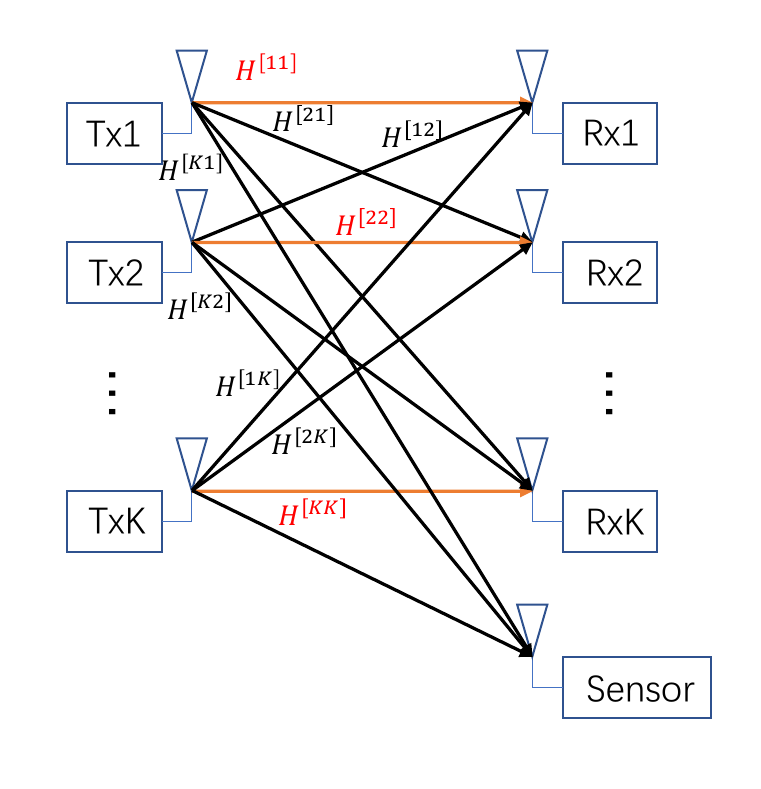}%
    }
    \subfloat[]{%
        \includegraphics[width=0.16\textwidth]{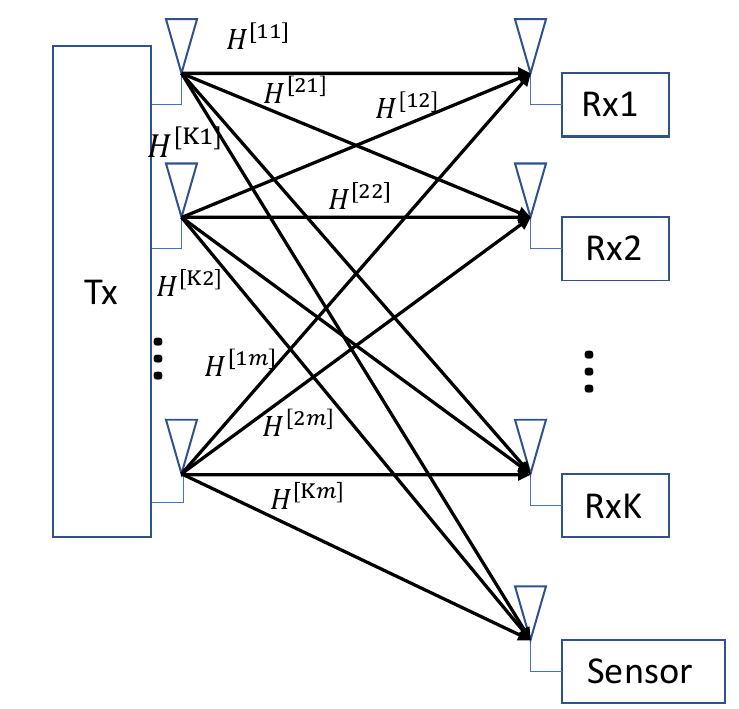}%
    }
    \subfloat[]{%
        \includegraphics[width=0.16\textwidth]{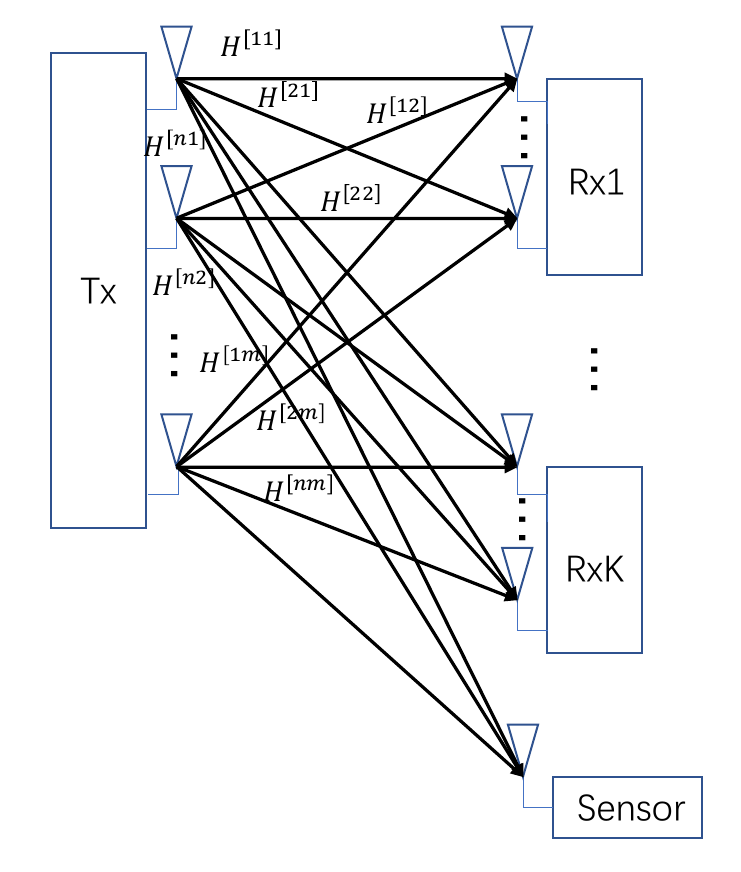}%
    }
    \caption{\small Heterogeneous coherence times model: (a) Interference channel, (b) MU-MISO channel, (c) MU-MIMO channel.}
    \label{system model}
\end{figure}

We investigate three different channels in this type:
\begin{itemize}
    \item {\it Interference channel.}  When $\Ksf_T = \Ksf_R = \Ksf$, $\msf_i=1$ for each  $i \in \mathcal{T}$, and $\nsf_j=1$  for each $j \in \mathcal{R}$, we obtain  the  interference channel as shown in Fig.~\ref{system model}(a). Note that in the interference channel, the message symbols requested by the $i$-th  receiver are only available at the $i$-th transmitter. 
    \item {\it MU-MISO channel.} When $\Ksf_T = 1$, $\Ksf_R = \Ksf$, $\msf_1=\msf$, and $\nsf_j=1$ for each $j \in \mathcal{R}$, we obtain  the MU-MISO channel as shown in Fig.~\ref{system model}(a). 
    \item {\it MU-MIMO channel.}  When $\Ksf_T = 1, \Ksf_R = \Ksf, \msf_1=\msf$, we obtain  the MU-MIMO channel as shown in Fig.~\ref{system model}(c).
\end{itemize}

For the MU-MISO and MU-MIMO channels, an additional assumption is that the CSI of receivers are known to the transmitters in a non-causal manner\footnote{There is a big body of literature in TDD/FDD massive MIMO systems\cite{xie2016unified,fan2017angle}, where downlink channels can be efficiently estimated from uplink pilot signals by leveraging channel reciprocity in spatial and angular domains.}, implying that the transmitters has prior knowledge of the entire block's channel states.

\subsubsection{Heterogeneous connectivity}
\label{topology channel}
Now we consider the partial connectivity for network topology. 
 Assume that the  receivers have perfect Channel State Information at the Receiver (CSIR) about their connected links and 
the sensor have no CSIR about the connected links.

 Both the receivers and sensor have channels with long coherence time,
 such that we assume that  the channels remain constant during a single transmission block (i.e., $t_0$ time slots).
Thus we can exploit the inherent properties of the network structure in a stable   network to enhance the network performance.

For the network topology, we consider two existing partial connection networks,  
the regular network in~\cite{bondy1976graph,zhang2006introduction, scheinerman2011fractional,yi2015topological} and the   neighboring antidotes network in~\cite{maleki2014index,jafar2012elements}. These two topologies are widely adopted in topological interference management due to their structured yet flexible connectivity patterns. The $(\Ksf, d)$-regular network, also known as the Wyner-type model, offers symmetric and localized interference, enabling tractable analysis and fundamental benchmarking of interference alignment schemes. On the other hand, the $(\Ksf, U, D)$ neighboring antidotes network allows for asymmetric and generalized connectivity, effectively modeling heterogeneous interference scenarios where each receiver is protected from its closest interferers. Together, these models facilitate a comprehensive understanding of interference structures in ISAC systems under limited CSI assumptions.

\begin{defn}[$(\Ksf, U, D)$ neighboring antidotes network \cite{maleki2014index,jafar2012elements}]
    \label{def:neighboring antidotes network}
    The $(\Ksf, U, D)$ neighboring antidotes network with $\Ksf>U+D$ and $D\geq U$, includes $\Ksf$ single-antenna transmitters and $\Ksf$ single-antenna users.  Each user $j$ is disconnected from the $U$ preceding transmitters (indexed $j-U, j-U+1, \dots, j-1$) and the $D$ succeeding transmitters (indexed $j+1, j+2, \dots, j+D$), where indices are interpreted modulo $\Ksf$.
       User $j$ connects to all remaining $\Ksf - U - D$ transmitters. Formally,  
      \( \mathcal{R}_j = \{ j, <j + D + 1>_{\Ksf}, \ldots, <j - U - 1>_{\Ksf} \} \).
  In addition,  exchanging the values of $U$ and $D$ results in an equivalent network.  For this topology, we consider {\it non-cooperative transmitters}, where each transmitter only has the communication messages desired by its corresponding receiver.
\end{defn}

\begin{defn}[$(\Ksf, d)$ regular network \cite{bondy1976graph,zhang2006introduction, scheinerman2011fractional,yi2015topological}]
    \label{def:regular network}
    The $(\Ksf, d)$-regular network  includes $\Ksf$ single-antenna transmitters and $\Ksf$ single-antenna users.  Each user $j$ receives signals from  transmitter $j$ as well as the successive $d-1$ ones, i.e., the set of connected transmitters by user $j$ is $\mathcal{R}_j = \{j, <j + 1>_{\Ksf}, \dots, <j + d - 1>_{\Ksf}\}$. This model with regular linear connectivity (cells are enumerated sequentially) is also known as Wyner-type model for cellular networks and has been widely investigated in information theory~\cite{wyner2002shannon,xu2011accuracy,simeone2012cooperative}.  For this topology, we consider {\it cooperative transmitters}, where each transmitter  has all the communication messages desired by the receivers.
\end{defn}

Based on the above   networks, we consider the following way to add the sensor into the network: In the $(\Ksf+1, U, D)$ neighboring antidotes network or the $(\Ksf+1, d)$ regular network, 
    we set the first $\Ksf$ users as the receivers and the last user as the sensor.
Define $\Rc^{[k]}_c$ and $\Rc_s$ as the set of connected transmitters by communication receiver $k$ and by the sensor, respectively.

\begin{figure}[htbp]
    \centering
    \subfloat[]{%
        \includegraphics[width=0.24\textwidth]{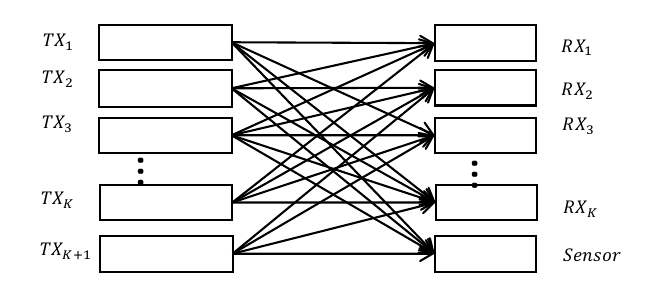}%
    }
    \subfloat[]{%
        \includegraphics[width=0.24\textwidth]{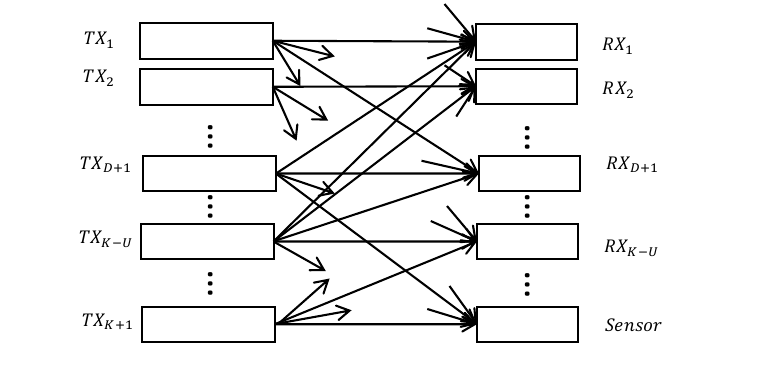}%
    }
    \caption{\small Heterogeneous connectivity model: (a) $(\Ksf+1, d)$-regular network, (b) $(\Ksf+1, U, D)$ neighboring antidotes network.}
    \label{TIM system model}
\end{figure}

\subsection{Performance Metrics} \label{matrics}

\subsubsection{Communication Degree of Freedom (cDoF)}
To evaluate communication performance, we define $\text{cDoF}$ as the sum degree of freedom per time slot, averaged over $t_0$ time slots. Each independent message symbol contributes one unit of communication DoF. The $\text{cDoF}$ is computed as the total number of  message symbols successfully decoded by all receivers, normalized by $t_0$. This metric quantifies the efficiency of communication in terms of the number of recoverable symbols over a given transmission duration.

\subsubsection{Sensing Degree of Freedom (sDoF)}
\label{def:sdof}

To facilitate the characterization of the fundamental tradeoff between communication and sensing, we adopt a dual metric to quantify both functionalities. Thus we consider the sensing degree of freedom based on the number of effective (i.e., independent) observations of the target during the estimation and detection process as in~\cite{xiong2023fundamental,xie2024sensing}.
More specifically, assuming the total number of transmitting antennas connected to the sensor is \(\msf'\), an effective observation sequence of length \(N\geq \msf'\) implies that the transmitters emits a set of orthogonal deterministic training sensing signals \(\xv_1^{[0]},\xv_2^{[0]},\ldots,\xv_N^{[0]}\) (each with dimension $\msf'\times 1$), satisfying 
\begin{align}
    \begin{bmatrix}
\xv_1^{[0]}, \xv_2^{[0]}, \ldots, \xv_N^{[0]}
\end{bmatrix}
\begin{bmatrix}
\xv_1^{[0]}, \xv_2^{[0]}, \ldots, \xv_N^{[0]}
\end{bmatrix}^H
=  \textbf{I}_{\msf'}.\label{eq:orthogonal}
\end{align}
Accordingly, as in~\cite{xiong2023fundamental,xie2024sensing} the total sensing degree of freedom (sDoF) is defined as the total number of the
independent observations $N$. Assume that the sensor obtains these $N$ independent observations by using $T$ time slots. The average sDoF (or just simply called sDoF in the rest) is $N/T$.\footnote{\label{foot:orthogonal} Note that when the statistic CSI is known at the transmitters, the transmitted deterministic sensing signal can be $\begin{bmatrix}
\xv_1^{[0]}, \xv_2^{[0]}, \ldots, \xv_N^{[0]}
\end{bmatrix} = \Wm\Sm_D$ with deterministic training signal $\Sm_D\in \mathbb{C}^{\msf'\times N}$ satisfying \(\Sm_D\Sm_D^H=\textbf{I}_{\msf'}\) and a precoding matrix \(\mathbf{W} \in \mathbb{C}^{\msf'\times \msf'}\) optimized to minimize detection/estimation errors by using the statistic CSI. Thus~\eqref{eq:orthogonal} becomes $\begin{bmatrix}
\xv_1^{[0]}, \xv_2^{[0]}, \ldots, \xv_N^{[0]}
\end{bmatrix}
\begin{bmatrix}
\xv_1^{[0]}, \xv_2^{[0]}, \ldots, \xv_N^{[0]}
\end{bmatrix}^H
= \Wm \Wm^H$. Except the modification on~\eqref{eq:orthogonal}, the definition of sensing DoF remains the same. The proposed ISAC schemes in this paper can be directly extended according to this definition of sDoF, while the achieved tradeoff points between cDoF and sDoF do not change.}

\textbf{Objective:} 
The objective of this paper is to characterize the capacity region of all achievable tradeoff points  $(\text{sDoF}, \text{cDoF})$, for large enough $P$ and some finite $t_0$.

\section{Main Results} 
\label{sec:main results}

 The main technical contribution  is to   introduce the
strategies of BIA and TIM  into the bistatic ISAC systems to completely
eliminate the interference from the communication messages
at the sensor, under the constraint that the sensor channel is
unknown to all. Since the dedicated sensing signals
are known to the  receivers while decoding the
desired messages, they can  eliminate the interference from the sensing signals. 
\begin{itemize}
    \item {\it Heterogeneous coherence times.} We can use the BIA strategy to eliminate the interference to the sensor. Meanwhile, 
    to eliminate the interference from the
undesired communication messages for each 
receiver, the zero forcing  (ZF) strategy is used.  
    \item {\it Heterogeneous connectivity.} Even though the CSI is unknown to the transmitters,  we can use the TIM strategy to align interference for all the receivers and  sensor by exploiting the symmetric network topology. 
\end{itemize}

\subsection{ISAC Model with Heterogeneous Coherence Times}

\begin{theorem}[Interference channel]
    \label{thm:IC region}
    For the bistatic ISAC systems with heterogeneous coherence times and  interference channel including $\Ksf$ single-antenna transmitters, $\Ksf$ single-antenna receivers, and one sensor, the lower convex  hull 
    of the following tradeoff points is achievable:
    \begin{align}
    (\text{sDoF}, \text{cDoF})=( (\Ksf-1)/\Ksf ,1), 
    \label{eq:ic point}
\end{align}
 $ (\text{sDoF}, \text{cDoF})=( 1 ,0)$, and $ (\text{sDoF}, \text{cDoF})=( 0 ,1)$, where the latter two points are achieved by the existing sensing-only scheme and   communication-only schemes.\footnote{\label{IA+BIA}Note that owing to the delay requirement of ISAC systems (especially for the sensing tasks), we consider zero forcing for communications within a limited number of coherence blocks, which excludes the vanilla interference alignment strategies that achieve precise interference alignment at the cost of an extremely large block length~\cite{jafar2008degrees}. Moreover, even compared with the time-sharing between $ (\text{sDoF}, \text{cDoF})=( 1 ,0)$ and  $ (\text{sDoF}, \text{cDoF})=( 0 ,\Ksf/2)$, when $\text{sDoF}=(\Ksf-1)/\Ksf$ our scheme can achieve $\text{cDoF}=1$, while the above time-sharing can only achieve $\text{cDoF}=1/2$.}

\end{theorem}

 Compared to the communication-only  tradeoff point $ (\text{sDoF}, \text{cDoF})=( 0 ,1)$ achieved by  ZF  (the optimal one-shot linear coding), in~\eqref{eq:ic point} we can maintain the same cDoF while additionally obtaining a sDoF equal to $(\Ksf-1)/\Ksf$. 
 
\begin{theorem}[MU-MISO channel]
    \label{thm:MUMISO}
    For the bistatic ISAC systems with  heterogeneous coherence times and the MU-MISO channel including one transmitter with $\msf$  antennas, $\Ksf$ single-antenna receivers, and one sensor, the lower convex hull of the following tradeoff points is achievable: 
    \begin{equation}
    (\text{sDoF}, \text{cDoF}) = \left(\frac{\left\lceil\frac{\msf}{\Ksf}\right\rceil-1}{\left\lceil\frac{\msf}{\Ksf}\right\rceil}, \frac{\msf}{\left\lceil\frac{\msf}{\Ksf}\right\rceil}\right),
\end{equation}
  $ (\text{sDoF}, \text{cDoF})=( 1 ,0)$, and  $ (\text{sDoF}, \text{cDoF})=( 0 ,\min\{\msf,\Ksf\} )$, where the latter two points are achieved by the existing sensing-only  and  communication-only schemes.
\end{theorem}

\begin{theorem}[MU-MIMO channel]
    \label{thm:MUMIMO}
    For the bistatic ISAC systems  with heterogeneous coherence times and the MU-MIMO channel including one transmitter with $\msf$  antennas, $\Ksf$ multi-antenna receivers (with the number of antennas  $\nsf_i, i=1,\ldots,\Ksf$), and one sensor, the lower convex hull of the following tradeoff points is achievable: 
    \begin{equation}
(\text{sDoF}, \text{cDoF}) = \left(\frac{\left\lceil\frac{\msf}{\sum_{k=1}^{\Ksf}\nsf_k}\right\rceil-1}{\left\lceil\frac{\msf}{\sum_{k=1}^{\Ksf}\nsf_k}\right\rceil}, \frac{\msf}{\left\lceil\frac{\msf}{\sum_{k=1}^{\Ksf}\nsf_k}\right\rceil}\right),
\end{equation}
 $ (\text{sDoF}, \text{cDoF})=( 1 ,0)$, and $ (\text{sDoF}, \text{cDoF})=( 0 ,\min\{\msf,\sum_{k=1}^{\Ksf}\nsf_k\} )$, where the latter two points are achieved by the existing sensing-only   and   communication-only schemes.
\end{theorem}
The description of the proposed ISAC schemes based on BIA for the interference channel and the MU-MISO channel can be found in Sections~\ref{sub:IA} and~\ref{proof:MUSiMO} respectively, while  the ISAC scheme based on BIA for the MU-MIMO channel  can be found in Appendix~\ref{pro:MUMIMO}.

\subsection{ISAC Model with Heterogeneous Connectivity}
\label{sub:connectivity}
For the bistatic ISAC model with heterogeneous connectivity described in Section~\ref{sub:channel model}, Theorems~\ref{thm:UDnco+1 region} and~\ref{thm:KDco+1 region} consider  the   two topology network  $(\Ksf+1, U, D)$ neighboring antidotes network and $(\Ksf+1, d)$ regular network , respectively.

\begin{theorem}[$(\Ksf+1, U, D)$ neighboring antidotes network]
    \label{thm:UDnco+1 region}
  For the bistatic ISAC systems with heterogeneous connectivity based on the $(\Ksf+1, U, D)$ neighboring antidotes network with $\Ksf+1>U+D$ and $D\geq U$,  the lower convex hull of the following trade-off points is achievable:  
    \begin{align}
        (\text{sDoF}, \text{cDoF}) = \left( \frac{U+1}{\Ksf-D+U+1}, \frac{\Ksf(U+1)}{\Ksf-D+U+1} \right),
        \label{eq:noncooperative point}
    \end{align}
      \((\text{sDoF}, \text{cDoF}) = (1, 0)\), and \((\text{sDoF}, \text{cDoF}) = \left( 0, \frac{\Ksf(U+1)}{\Ksf-D+U+1} \right)\), where the latter two points are achieved by the existing sensing-only scheme and the communication-only scheme \cite{maleki2014index, jafar2012elements}; ,  
\end{theorem}

\begin{theorem}[\( (\Ksf+1, d) \)  regular network]
    \label{thm:KDco+1 region}
  For the bistatic ISAC systems with heterogeneous connectivity and  cooperative transmitters,  based on the \( (\Ksf+1, d) \)  regular network, 
         the lower convex hull of the following trade-off points are achievable: 
    \begin{equation}
    (\text{sDoF}, \text{cDoF}) = \left( \frac{2}{d+1}, \frac{2\Ksf}{d+1} \right),
    \label{eq:coo topo1}
    \end{equation}
      \((\text{sDoF}, \text{cDoF}) = (1, 0)\), and \((\text{sDoF}, \text{cDoF}) = \left( 0, \frac{2\Ksf}{d+1} \right)\), where the latter two points are achieved by the existing sensing-only  and communication-only schemes~\cite{yi2015topological}.

\end{theorem}
The description of the proposed ISAC schemes based on TIM for the $(\Ksf+1, U, D)$ neighboring antidotes network and the $(\Ksf+1, d)$ regular network can be found in Sections~\ref{sub:neighboring} and~\ref{sub:regular network}, respectively. 

\begin{remark}[Extended framework for the proposed schemes: Adding a sensor]
\label{rem:extension}
Given an existing topological network, 
another possible way  to consider the sensor into the network is to purely add the sensor into the network, instead of replacing one receiver by this sensor as we considered previously. More precisely, in the $(\Ksf, d)$ regular network or the $(\Ksf, U, D)$ neighboring antidotes network,  with the original $\Ksf$ users being receivers and one additional user as the sensor, where the set of connected transmitters of the sensor is a subset of one receiver.

The proposed ISAC schemes for the $(\Ksf+1, d)$ regular network or the $(\Ksf+1, U, D)$ neighboring antidotes network could be directly extended to this framework. 
For  the $(\Ksf, U, D)$ neighboring antidotes network with an adding sensor,  the lower convex hull of the following trade-off points is achievable:
 \begin{equation}
        (\text{sDoF}, \text{cDoF}) = \left( \frac{U+1}{\Ksf-D+U}, \frac{\Ksf(U+1)}{\Ksf-D+U} \right),
        \label{eq:extension KUD}
    \end{equation}
     \((\text{sDoF}, \text{cDoF}) = (1, 0)\), and \((\text{sDoF}, \text{cDoF}) = \left( 0, \frac{\Ksf(U+1)}{\Ksf-D+U} \right)\), where the latter two points are achieved by the existing sensing-only scheme and communication-only schemes~\cite{maleki2014index, jafar2012elements};
     the detailed description of the scheme  can be found in the  Appendix~\ref{pro:UDnco+1To2}.
     
     For   the $(\Ksf, d)$ regular network with an adding sensor, the lower convex hull of the following trade-off points is achievable:
\begin{align}
    (\text{sDoF}, \text{cDoF}) = \left( \frac{1}{d+1}, \frac{2\Ksf}{d+1} \right),
    \label{eq:coo topo2}
    \end{align}
      \((\text{sDoF}, \text{cDoF}) = (1, 0)\), and \((\text{sDoF}, \text{cDoF}) = \left( 0, \frac{2\Ksf}{d+1} \right)\), where the latter two points are achieved by the existing sensing-only  and  communication-only schemes~\cite{yi2015topological}; the detailed description of the scheme  can be found in the  Appendix~\ref{pro:KDco+1}.
\end{remark}

\section{Achievable schemes for the ISAC Model with Heterogeneous Coherence Times}
This section introduces the proposed BIA-based ISAC schemes for bistatic ISAC systems under heterogeneous coherence times, covering interference channel, MU-MISO, and MU-MIMO channels. The approach evolves from single-antenna to multi-antenna configurations, harnessing system diversity to effectively incorporate sensing capabilities at the sensor side.


\subsection{Proof of Theorem~\ref{thm:IC region}: Interference Channel}
\label{sub:IA}

\begin{figure}
    \centering
    \includegraphics[width=0.6\linewidth]{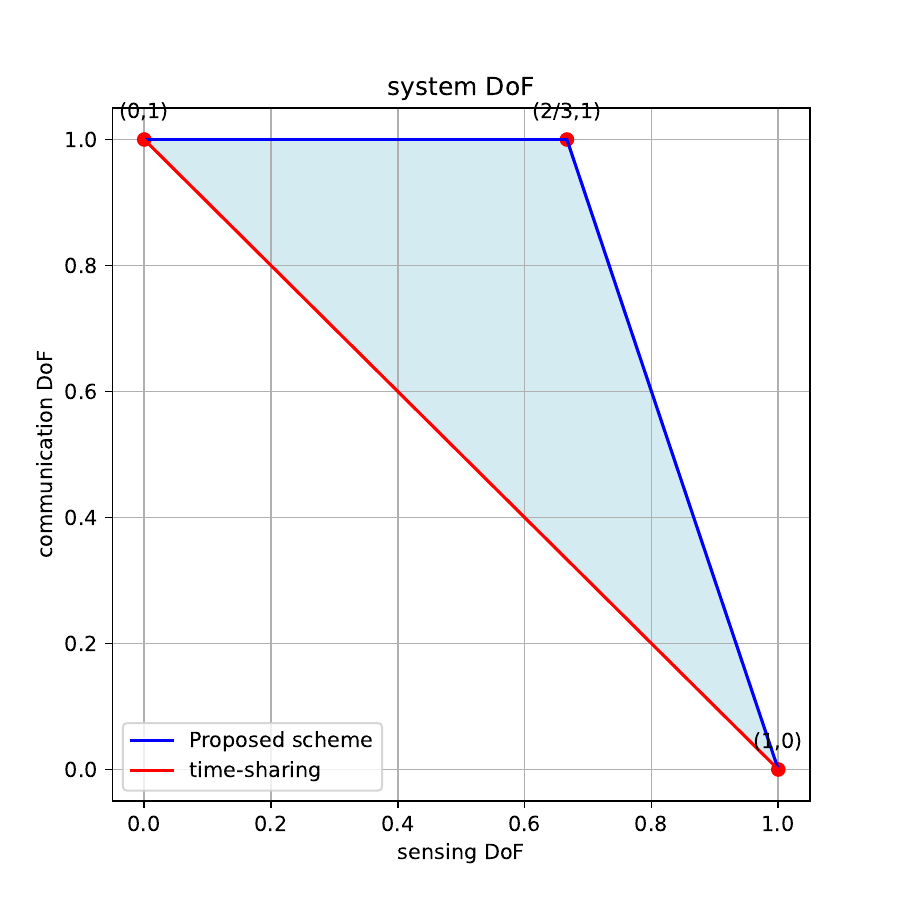}
    \caption{\small Tradeoff between sDoF and  cDoF for the $3\times 3$ interference channel  with a sensor.}
    \label{fig:34ICsystem dof}
\end{figure}

We first describe the main idea of the proposed scheme based on the BIA strategy through one example for the interference channel. 
\begin{example}[Interference channel with heterogeneous coherent times]
\label{subsub:BIA IC case}
Consider the bistatic ISAC system with heterogeneous coherence times and interference channel including 
$3$ single-antenna transmitters, $3$ single-antenna  receivers, and one sensor, as illustrated in Fig.~\ref{fig:senior34inter}.
For this system, the tradeoff points $ (\text{sDoF}, \text{cDoF})=( 1 ,0)$, $ (\text{sDoF}, \text{cDoF})=( 0 ,1 )$  can be achieved by the  sensing-only scheme and communication-only scheme, respectively. We then propose an ISAC scheme combining BIA and zero-forcing, achieving 
$ (\text{sDoF}, \text{cDoF})=(2/3, 1)$. Let $t_0=6$, and  
let the communication signals in the  $6$ time slots by each transmitter $k\in \{1,2,3\} $ be $x^{[k]}_c(1)=x^{[k]}_c(2)=x^{[k]}_c(3)=W^{[k]}_1$  and $x^{[k]}_c(4)=x^{[k]}_c(5)=x^{[k]}_c(6)=W^{[k]}_2$, 
where each $W^{[k]}_i\in \CC$ represents a message symbol (as defined in Footnote~\ref{foot:Gaussian coding}) desired by the \( k \)-th  receiver. 

Thus by removing the dedicated sensing signals (which are deterministic), 
  the \(k \)-th receiver obtains  from time slots $\{1,2,3\}$ :
\begin{align}
&    \hat{\yv}_1^{[k]} = 
    \begin{bmatrix} 
        \hat{y}^{[k]}(1) \\ 
        \hat{y}^{[k]}(2) \\ 
        \hat{y}^{[k]}(3) 
    \end{bmatrix} 
    = \sum_{i=1}^{3}\begin{bmatrix} 
        H^{[ki]}(1)x^{[k]}_c(1) \\ 
        H^{[ki]}(2)x^{[k]}_c(2) \\ 
        H^{[ki]}(3)x^{[k]}_c(3)
    \end{bmatrix}  
    + \begin{bmatrix} 
        z^{[k]}(1) \\ 
        z^{[k]}(2) \\ 
        z^{[k]}(3)
    \end{bmatrix}\nonumber \\ 
    &= \sum_{i=1}^{3}\begin{bmatrix} 
        H^{[ki]}(1) \\ 
        H^{[ki]}(2) \\ 
        H^{[ki]}(3)
    \end{bmatrix} W_1^{[i]}
    + \begin{bmatrix} 
        z^{[k]}(1) \\ 
        z^{[k]}(2) \\ 
        z^{[k]}(3)
    \end{bmatrix}.
\end{align}
\begin{figure}
    \centering
    \includegraphics[width=0.8\linewidth]{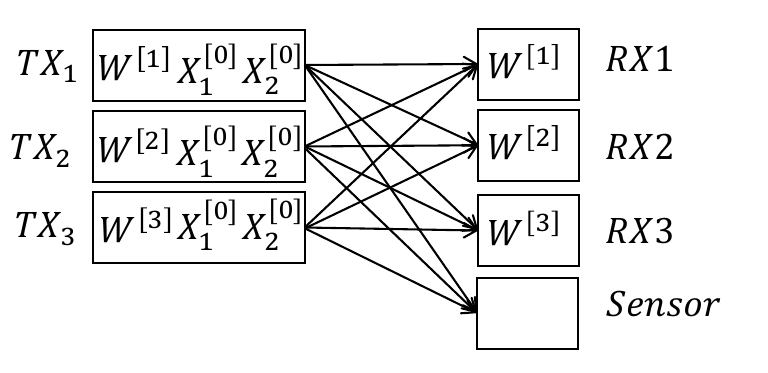}
    \caption{\small  $3\times 3$ interference channel with a sensor for Example~1.}
    \label{fig:senior34inter}
\end{figure}

For the \( k \)-th  receiver, since the channel coefficients $H^{[ki]}(t)$ where $t, i\in \{1,2,3\}$ are i.i.d. with circularly symmetric complex Gaussian  distribution, 
we  can apply the zero-forcing decoding to decode $W_1^{[k]}$ from $\hat{\yv}_1^{[k]}$. 
Similarly, the \( k \)-th  receiver can recover $W^{[k]}_2$ from time slots $\{4,5,6\}$.  
Thus the achieved communication DoF is $1$.

    The sensing DoF  is $2/3$, by obtaining $4$ effective observations 
    can be obtained over $6$ time slots. Next, we demonstrate the signal design using blind interference alignment. Let \(\xv_k^{[0]} = [x_{k,1}, x_{k,2}, x_{k,3}]^T \in \mathbb{C}^{3 \times 1}\) for \(k = 1, 2, 3, 4\), satisfying \( [\xv_1^{[0]}, \xv_2^{[0]},\xv_3^{[0]},\xv_4^{[0]}][\xv_1^{[0]}, \xv_2^{[0]},\xv_3^{[0]},\xv_4^{[0]}]^H = \mathbf{I}_{3} \). 
    Our idea is to use the fact that the channel for the sensor remains constant over a certain period and that the communication signals also remain constant to cancel the communication signals, which are interference for the sensor. 
For the \(k \)-th transmitter where $k\in \{1,2,3\}$, the transmitted signals over  time slots $\{1,2,3\}$ are: \( x^{[k]}(1)= W^{[k]} + x_{1,k} + x_{2,k} \), \( x^{[k]}(2) = W^{[k]} + x_{2,k} \), \( x^{[k]}(3)=W^{[k]} + x_{1,k} \). The signals received by the sensor can be expressed as:
\begin{align*}
&\yv^{[4]}_{s,1}= 
 \begin{bmatrix} 
       y^{[4]}_s (1) \\ 
      y^{[4]}_s (2) \\ 
       y^{[4]}_s (3)
    \end{bmatrix} 
 \\&   =
\sum_{i=1}^{3}\begin{bmatrix} 
        H^{[4i]}(1)(W^{[i]}+x_{1,i} + x_{2,i})\\ 
        H^{[4i]}(1)(W^{[i]}+x_{2,i})\\ 
        H^{[4i]}(1)(W^{[i]}+x_{1,i})
    \end{bmatrix}
    +\begin{bmatrix} 
        z_s(1) \\ 
        z_s(2) \\ 
        z_s(3)
    \end{bmatrix}.
\end{align*}
By subtracting the observations at the second and third time slots from the first time slot, we get:
\begin{align*}
    y_{s}^{[4]}(1) - y_{s}^{[4]}(t)
    &= \sum_{i=1}^{3}H^{[4i]}(1) x_{t-1,i} + z_s(1)-z_s(t) \\
    &= \hv\begin{bmatrix} 
        x_{t-1,1} \\ 
        x_{t-1,2} \\ 
        x_{t-1,3} 
    \end{bmatrix}+z_s(1)-z_s(t), 
\end{align*} 
where \( \hv = \begin{bmatrix} H^{[41]}(1), H^{[42]}(1), H^{[43]}(1) \end{bmatrix} \) and  \(t={2,3}\). Therefore, the sensor  obtains \( \hv \xv_1^{[0]} + z_s(1)-z_s(2) \) and \( \hv \xv_2^{[0]} +z_s(1)-z_s(3) \), meaning two effective observations through $\xv_1^{[0]}, \xv_2^{[0]}$ over the first three time slots. Similarly, the sensor obtains another two  effective observations through $\xv_3^{[0]}, \xv_4^{[0]}$ over  time slots $\{4,5,6\}$. 
Thus the sDoF is $2/3$, and the achieved tradeoff point by the proposed scheme is  
$ (\text{sDoF}, \text{cDoF})=(2/3, 1)$.
Considering the three achieved tradeoff points, the lower convex hull is plotted by the blue line in Fig.~\ref{fig:34ICsystem dof}, while the red line represents the time-sharing between the  sensing-only and communication-only points. It can be seen that the achieved point by the proposed scheme significantly improves the one by time-sharing.

\end{example}

We are now ready to generalize the above example  achieving the tradeoff point $(\text{sDoF}, \text{cDoF}) = \left((\Ksf-1)/\Ksf, 1\right)$, for the  \(\Ksf \times \Ksf\) interference channel with an additional sensor. 

Let $t_0=2\Ksf$, and  
let the communication signals in the first $\Ksf$ time slots by each transmitter $k\in \{1,\ldots,\Ksf\} $ be \(x^{[k]}_c(1)=\cdots=x^{[k]}_c(\Ksf)=W_1^{[k]}\), 
and the communication signals in the second $\Ksf$ time slots by each transmitter $k$ be \(x^{[k]}_c(\Ksf+1)=\cdots=x^{[k]}_c(2\Ksf)=W_2^{[k]}\), 
where each $W_i^{[k]}\in \CC$ represents a message symbol  desired by the \( k \)-th  receiver. 

At  time slot $t$, the signal received by the \(k\)-th receiver is, 
\begin{align*}
    &y^{[k]}(t) = H^{[k1]}(t) (x_c^{[1]}(t) + x_s^{[1]}(t)) + H^{[k2]}(t) (x_c^{[2]}(t) \\
   & + x_s^{[2]}(t)) + \dots + H^{[k\Ksf]}(t) (x_c^{[\Ksf]}(t) + x_s^{[\Ksf]}(t)) + z^{[k]}(t),
\end{align*}
for each $t\in \{1,\ldots,2\Ksf\}$. 
Since the CSI and the transmitted sensing signal \(x_s^{[k]}(t) \) are known to the receivers, we can cancel this part of the signal at the receiver, 
resulting in an estimate \( \hat{\yv}^{[k]} \) that only concerns the communication signals in $\Ksf$ time slots:
\begin{align}
&    \hat{\yv}_1^{[k]} = 
    \begin{bmatrix} 
        \hat{y}^{[k]}(1) \\ 
        \hat{y}^{[k]}(2) \\
        \vdots \\
        \hat{y}^{[k]}(\Ksf) 
    \end{bmatrix} 
    = \sum_{i=1}^{\Ksf}\begin{bmatrix} 
        H^{[ki]}(1)x^{[k]}_c(1) \\ 
        H^{[ki]}(2)x^{[k]}_c(2) \\ 
        \vdots \\
        H^{[ki]}(\Ksf)x^{[k]}_c(\Ksf)
    \end{bmatrix}  
    + \begin{bmatrix} 
        z^{[k]}(1) \\ 
        z^{[k]}(2) \\
        \vdots \\
        z^{[k]}(\Ksf)
    \end{bmatrix}\nonumber \\ 
    &= \sum_{i=1}^{\Ksf}\begin{bmatrix} 
        H^{[ki]}(1) \\ 
        H^{[ki]}(2) \\ 
        \vdots \\
        H^{[ki]}(\Ksf)
    \end{bmatrix} W_1^{[i]}
    + \begin{bmatrix} 
        z^{[k]}(1) \\ 
        z^{[k]}(2) \\ 
        \vdots \\
        z^{[k]}(\Ksf)
    \end{bmatrix}.
\end{align}
Since the channel coefficients $H^{[ki]}(t)$ where $t, i\in \{1,\ldots,\Ksf\}$ are i.i.d. with circularly symmetric complex Gaussian  distribution, the $k$-th receiver can decode $W_1^{[k]}$ from the first $\Ksf$ time slots with high probability by using ZF. 
Similarly, it can decode  $W_2^{[k]}$ from the second $\Ksf$ time slots.
Thus the cDoF is \( 1\).


Over \(2\Ksf\) time slots, the sensor obtains \(2\Ksf - 2\) effective observations by transmitting \(2\Ksf - 2\) sensing signals 
\(\xv_k^{[0]} = [x_{k,1}, \ldots, x_{k,\Ksf}]^T \in \mathbb{C}^{\Ksf \times 1}\), where \(k = 1, 2, \ldots, 2\Ksf - 2\), satisfying \([\xv_1^{[0]}, \xv_2^{[0]}, \dots, \xv_{2\Ksf-2}^{[0]}] [\xv_1^{[0]}, \xv_2^{[0]}, \dots, \xv_{2\Ksf-2}^{[0]}]^H=  \textbf{I}_\Ksf\). 

Let us first consider the first $\Ksf$ time slots. We  design the transmitted dedicated sensing signals 
$x_s^{[k]}(t)$ for $t\in \{1,\ldots,\Ksf\}$ as follows, 

\begin{equation}
    \begin{aligned}
  \begin{bmatrix}
       x^{[k]}_s(1) \\
       x^{[k]}_s(2) \\
        \vdots \\
        x^{[k]}_s(\Ksf)
    \end{bmatrix}    =
    \begin{bmatrix}
        1 & 1 & 1 & \dots & 1 \\
        0 & 1 & 1 & \dots & 1 \\
        1 & 0 & 1 & \dots & 1 \\
        \vdots & \vdots & \vdots & \ddots & \vdots \\
        1 & 1 & 1 & \dots & 0
    \end{bmatrix}
    \begin{bmatrix}
        x_{1,k} \\
        x_{2,k} \\
        x_{3,k} \\
        \vdots \\
        x_{\Ksf-1,k}
    \end{bmatrix}.
    \end{aligned}
\end{equation}
Recall that the communication message transmitted by each transmitter during the first $\Ksf$ time slots remains the same, as well as the channel from each transmitter to the sensor. 
Thus for each $\tau\in \{2,3,\ldots,\Ksf\}$, we have
\begin{align*}
   &\Delta y_s^{[\Ksf+1]}(\tau) = y_s^{[\Ksf+1]}(1) - y_s^{[\Ksf+1]}(\tau) = H^{[\Ksf+1,1]}(1) x_{\tau-1,1}  +\\  &  
    H^{[\Ksf+1,2]}(1) x_{\tau-1,2}+ \cdots + H^{[\Ksf+1,\Ksf]}(1) x_{\tau-1,\Ksf}+z_s(1)-z_s(\tau)\\ 
    &=  \begin{bmatrix}
H^{[\Ksf+1,1]}(1),  \dots, H^{[\Ksf+1,\Ksf]}(1)
\end{bmatrix}
\xv_{\tau-1}^{[0]} +z_s(1)-z_s(\tau).
\label{eq:differ}
\end{align*} 

Thus over the first \(\Ksf\) time slots, we obtain \(\Ksf-1\) effective observations through $\xv_1^{[0]},\ldots,\xv_{\Ksf-1}^{[0]}$. Similarly, over the second \(\Ksf\) time slots, we can obtain another \(\Ksf-1\) effective observations through $\xv_{\Ksf}^{[0]},\ldots,\xv_{2\Ksf-2}^{[0]}$. Hence, the sDoF is \(\frac{\Ksf-1}{\Ksf}\).

\subsection{Proof of Theorem~\ref{thm:MUMISO}: MU-MISO}
\label{proof:MUSiMO}
Now we consider the bistatic ISAC systems with heterogeneous coherent times and the MU-MISO channel including one transmitter with $\msf$  antennas, $\Ksf$ single-antenna communication receivers, and one sensor.
By Theorem~\ref{thm:MUMISO}, we propose a new BIA scheme 
to achieve $(\text{sDoF}, \text{cDoF}) = \left(\frac{\left\lceil\frac{\msf}{\Ksf}\right\rceil-1}{\left\lceil\frac{\msf}{\Ksf}\right\rceil}, \frac{\msf}{\left\lceil\frac{\msf}{\Ksf}\right\rceil}\right)$.  
Note that when $\msf \leq \Ksf$,  the above tradeoff point becomes $(0,\msf)$, which can be simply achieved by the  communication-only scheme. Hence, in this proof we only consider the case $\msf>\Ksf$. 
For the ease of notation, we define that \(\lceil \msf/\Ksf \rceil :=\asf \), thus $\asf>1$; define that $\psf = <\msf>_{\Ksf}$.  

Let $t_0= \asf\left\lceil\frac{\msf}{\asf-1}\right\rceil $.
During these $t_0$ time slots,  the sensor 
 will obtain \(\left\lceil\frac{\msf}{\asf-1}\right\rceil(\asf-1)\) effective observations,  through the sensing signals $\xv_1^{[0]}, \ldots, \xv_{\left\lceil\frac{\msf}{\asf-1}\right\rceil(\asf-1)}^{[0]} \in \mathbb{C}^{\msf\times 1}$, satisfying \(\left[\xv_1^{[0]}, \ldots, \xv_{\left\lceil\frac{\msf}{\asf-1}\right\rceil(\asf-1)}^{[0]}\right] \left[\xv_1^{[0]}, \ldots, \xv_{\left\lceil\frac{\msf}{\asf-1}\right\rceil(\asf-1)}^{[0]}\right]^H= I\).  In each period of $\asf$ time slots, we let  the receivers totally recover $\msf$ communication messages, and let the sensor 
obtain $\asf-1$ effective observation.  

In the following, we illustrate our proposed scheme for the first  $\asf$ time slots. Each receiver $k\in \{1,\ldots,\psf\} $ should  decode the communication messages $W_{j}^{[i]} \in \CC$ where $j\in \{1,2,...,\asf\}$, and each receiver $k\in \{\psf+1,\ldots,\Ksf\} $ should decode the communication messages $W_{j}^{[i]} \in \CC$ where $j \in \{1,2,3,...,\asf-1\}$.

For each time slot $t\in \{1,\ldots,\asf\}$, we design  the communication signals by the transmitter with $\msf$ antennas as:
\begin{equation*}
 \xv_c(1)=\cdots=   \xv_c(\asf) = \sum^{\asf-1}_{\tau=1}\sum^{\Ksf}_{k=1}\vv^{[k]}_{\tau} W^{[k]}_{\tau} + \sum^{\psf}_{j=1}\vv^{[j]}_{\asf} W^{[j]}_{\asf} \in\CC^{\msf \times 1},
\end{equation*}
where $ \{\vv^{[k]}_{t} : k\in \{1,\ldots,\Ksf\}, t\in \{1,\ldots,\asf-1\}\} $ and $\{\vv^{[j]}_{\asf}:j\in \{1,\ldots,\psf\} \}$  are the precoding vectors to be determined later, each with dimension $\msf \times 1$. 
At time slot $t\in \{1,\ldots,\asf\}$, for each $k\in\{1,\ldots,\Ksf\}$, the signal received by the $k$-th receiver is, 
\begin{align*}
    y^{[k]}(t) &=  [h^{[k1]}(t) ,   \ldots,   h^{[k\msf]}(t)]  (\xv_c(t)+\xv_s(t)) +z^{[k]}(t), 
\end{align*}
where $h^{[ki]}(t)$ represents the channel coefficient from $i$-th antenna of the transmitter to $k$-th receiver at time slot $t$, for each $k\in \{1,\ldots,\Ksf\}$ and $t\in \{1,\ldots,\asf\}$. By removing the dedicated signals, the $k$-th  receiver obtains,
\begin{align*}
    \hat{y}^{[k]}(t) &= [ h^{[k1]}(t) ,  \dots , h^{[k\msf]}(t)] \xv_c(t)+z^{[k]}(t). 
\end{align*} 

The precoding vectors are designed to ensure that 
\begin{itemize}
    \item[c1.] for each $k\in \{1,\ldots,\psf\}$ and $t\in \{1,\ldots,\asf\}$, the $k$-th  receiver can decode $W^{[k]}_t$ from $\hat{y}^{[k]}(t)$;
    \item[c2.] for each $k\in \{\psf+1,\ldots,\Ksf\}$ and $t\in \{1,\ldots,\asf-1\}$, the $k$-th  receiver can decode $W^{[k]}_t$ from $\hat{y}^{[k]}(t)$. 
\end{itemize}
To satisfy the above conditions, we propose the following design on the precoding vectors:
\begin{itemize}
    \item Design on $\vv^{[k]}_{t}$ for $k\in \{1,\ldots,\psf\}, t \in \{1,\ldots,\asf\}$. We let $\vv^{[k]}_{t}$ be a right null vector of the following matrix with dimension $(\msf-1) \times \msf$:
\begin{equation}
\label{mat:MUMISO}
        \Hm_{t}^{[k]} = \begin{bmatrix}
    \Hm(1) \\
    \Hm(2) \\
    \vdots \\
    \Hm(t) \backslash \Hm(t)[k,:] \\
    \vdots \\
    \Hm\left(\asf\right)
\end{bmatrix},
\end{equation}
where  $ \Hm(i)$ is the channel matrix from the transmitter to the $\Ksf$ receivers at time slot \(i\in\ \{1,2,...,\asf-1\}\), 
$$ \Hm(i)=   \begin{bmatrix}
        h^{[11]}\left(i\right) & h^{[12]}\left(i\right) & \cdots & h^{[1\msf]}\left(i\right) \\
        h^{[21]}\left(i\right) & h^{[22]}\left(i\right) & \cdots & h^{[2\msf]}\left(i\right) \\
        \vdots & \vdots & \ddots & \vdots \\
        h^{[\Ksf1]}\left(i\right) & h^{[\Ksf2]}\left(i\right) & \cdots & h^{[\Ksf\msf]}\left(i\right)
    \end{bmatrix} \in \mathbb{C}^{\Ksf \times \msf},$$
    \( \Hm(i)[j, :] \) represents the \(j\)-th row of \( \Hm(i) \),  \( \Hm(i) \backslash \Hm(i)[j, :] \) represents the matrix \( \Hm(i) \) with the \(j\)-th row removed,  
    $$\Hm(\asf)= \begin{bmatrix}
        h^{[11]}\left(\asf\right) & h^{[12]}\left(\asf\right) & \cdots & h^{[1\msf]}\left(\asf\right) \\
        h^{[21]}\left(\asf\right) & h^{[22]}\left(\asf\right) & \cdots & h^{[2\msf]}\left(\asf\right) \\
        \vdots & \vdots & \ddots & \vdots \\
        h^{[\psf1]}\left(\asf\right) & h^{[\psf2]}\left(\asf\right) & \cdots & h^{[\psf\msf]}\left(\asf\right)
    \end{bmatrix} \in \mathbb{C}^{\psf \times \msf}$$ is the channel matrix from the transmitter to the first $\psf$ receivers at time slot $\asf$.

Note that the matrix $\Hm_{t}^{[k]}$ with dimension $(\msf-1)\times\msf$ has one non-zero right null vector with high probability.
\item  Design on $\vv^{[k]}_{t}$ for $k\in \{\psf+1,\ldots,\asf\}, t \in \{1,\ldots,\asf-1\}$.
Let $\vv_{t}^{[k]}$ be the right null vector of  $\Hm_{t}^{[k]}$, 
with the same definition as~\eqref{mat:MUMISO}. 
\end{itemize}

By the above selection on the precoding vectors, one can check that the decodability conditions c1 and c2 are satisfied with high probability. For example, let us focus on receiver $1$. Its received signal after removing the sensing signal at time slot $t\in \{1,\ldots,\asf\}$ is  
\begin{align}
    &\hat{y}^{[1]}(t) = \Hm(t)[1,:]\xv_c(t) +z^{[1]}(t) \nonumber\\
    &=\Hm(t)[1,:](\sum^{\asf-1}_{\tau=1}\sum^{\Ksf}_{k=1}\vv^{[k]}_{\tau} W^{[k]}_{\tau} + \negmedspace\negmedspace \sum^{\psf}_{j=1} \negmedspace  \vv^{[j]}_{\asf} W^{[j]}_{\asf} )+z^{[1]}(t). \label{eq:y1}
\end{align}
It can be checked that the product of $\Hm(t)[1,:]$ and each precoding vector in~\eqref{eq:y1} (except $\vv^{[1]}_{t}$) is $0$. Hence, receiver~$1$ can recover $W^{[1]}_{t}$ with high probability.

Thus the cDoF of the system is $\frac{\psf\asf+(\Ksf-\psf)(\asf-1)}{\asf} = \frac{m}{\asf}$.

We then  design the transmitted dedicated sensing signals $\xv_s(t)$ for $t\in \{1,\ldots,\asf\}$  as follows (recall that $\xv_1^{[0]}, \ldots, \xv_{\asf}^{[0]}$ have been selected before)
\begin{align*}
&\xv_s(1)=\sum^{\asf-1}_{i=1} \xv_i^{[0]}, \ \  \xv_s(t)=\sum^{\asf-1}_{i=1,i \neq t-1} \xv_i^{[0]}, \ \forall t\in \{2,\ldots,\asf\}.
\end{align*} 
At each time slot $t\in \{1,\ldots,\asf\}$, the sensor receives 
\begin{equation*}
    y_s^{[\Ksf+1]}(t) = \hv^{[\Ksf+1]}(t) (\xv_c(t)+\xv_s(t)) +z_s(t),
\end{equation*}
where \(\hv^{[\Ksf+1]}(t) = [ h^{[\Ksf+1,1]}(t) , \dots , h^{[\Ksf+1,\msf]}(t) ]\), and $h^{[\Ksf+1,i]}(t)$ represents  the channel coefficient from $i$-th antenna of the transmitter to the sensor. 

Since $\hv^{[\Ksf+1]}(1)=\cdots=\hv^{[\Ksf+1]}(\asf)$ and $\xv_c(1)=\cdots=\xv_c(\asf)$,
 the sensor can subtract the received signal at time slot \(t\in \{2,\ldots,\asf\}\) from the received signal at the first time slot:
\begin{align*}
    \yv_s^{[\Ksf+1]}(1) - \yv_s^{[\Ksf+1]}(t) 
    = \hv^{[\Ksf+1]}(1)\xv_{t-1}^{[0]}+z_s(1)-z_s(t),
\end{align*}
thereby canceling the interference caused by the communication signals and obtaining an effective observation of \(\xv_{t-1}^{[0]}\). Over \(\asf\) time slots, a total of \(\asf-1\) effective observations can be made, and the sensing degrees of freedom (SDoF) is \((\asf-1)/\asf\).

\begin{remark}[Connection of the proposed ISAC schemes based on BIA]
In this section, we present a new joint precoding framework tailored for ISAC systems, which explicitly exploits the heterogeneity in coherence times between receivers and sensor. 
 The key characteristics of the three proposed schemes (for the interference channel, MU-MISO and MU-MIMO channels) can be summarized as follows:
 The key characteristics of the schemes can be summarized as follows:
\begin{itemize}
\item 
The common point of the three proposed schemes 
is to leverage the ZF precoding to eliminate interference among receivers (with the  CSI of the communication channel) and the BIA strategy  to eliminate the interference from the communication messages to the sensor (without  CSI of the sensor channel).
  
     \item The primary distinction among the three  schemes arises from the antenna configurations and their implications on interference management. In the interference channel  with single-antenna transmitters and receivers, spatial beamforming is infeasible due to the insufficient antenna dimension. 
    Conversely, when the transmitter is equipped with multiple antennas, spatial beamforming gains can be exploited, providing additional degrees of freedom.
   Both MU-MISO and MU-MIMO scenarios share a common design principle: spatial beamforming combined with repeated symbol patterns across time slots creates structured observations that facilitate the separation of communication and sensing signals. 
\end{itemize}
\end{remark}

\section{Achievable schemes for the ISAC Model with Heterogeneous Connectivity}
\label{subsec:TIMscheme}
While previous designs rely on heterogeneous coherence times to separate sensing and communication signals, 
in this section we consider heterogeneous connectivity
 in infrastructures like cellular networks. Assume that the network topology (i.e., the user-to-base station connectivity) remains relatively stable over time, but the exact channel state information is unknown.
 As stated in Section~\ref{sub:connectivity}, we will consider two topologies,   $(\Ksf, U, D)$ neighboring antidotes network and   the $(\Ksf, d)$-regular network.

\begin{figure}
    \centering
    \includegraphics[width=1\linewidth]{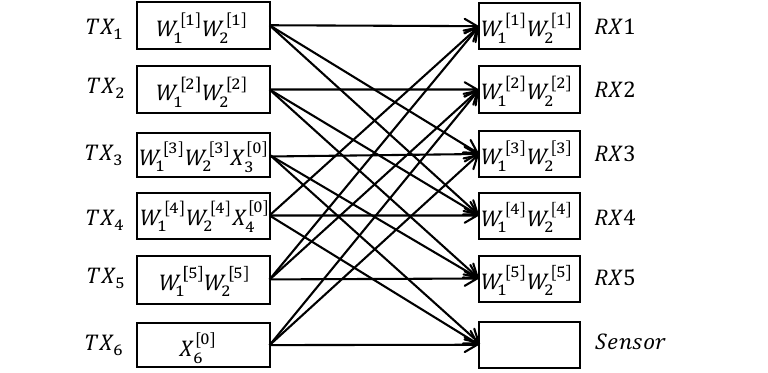}
    \caption{\small TIM non-cooperative interference channel in Example~2. 
    }
    \label{fig:noco6TIM}
\end{figure}

\subsection{Proof of Theorem~\ref{thm:UDnco+1 region}: $(\Ksf+1, U, D)$ neighboring antidotes network}
\label{sub:neighboring}
Considering the bistatic
ISAC systems with heterogeneous connectivity modeled as a $(\Ksf+1, U, D)$ neighboring antidotes network,  
we first  describe the main idea of the proposed TIM scheme through the following example. In the  bistatic
ISAC systems with heterogeneous connectivity,   all channels do not change during $t_0$ time slots; so for the ease of notation, we remove the time indices from channel coefficients. 
\begin{example}[ISAC system  with $(6, 1, 2)$ neighboring antidotes network]
\label{subsub:TIM noco 66}
Consider the bistatic ISAC system with a partially connected interference channel including $6$ single-antenna transmitters, $5$ single-antenna  receivers and one sensor, as illustrated in Fig.~\ref{fig:noco6TIM}.
Each receiver or sensor is disconnected from the previous \( U=1 \) and next \( D=2 \) transmitters, and connected to other $3$ transmitters; 
the topology belongs to  the $(6,1,2)$-neighboring antidotes network described  in Definition~\ref{def:neighboring antidotes network}.  We have 
\begin{align*}
    &\Rc^{[1]}_c= \{1,4,5\}, \ \Rc^{[2]}_c= \{2,5,6\}, \ \Rc^{[3]}_c= \{1,3,6\}, \\
    &\Rc^{[4]}_c= \{1,2,4\}, \ \Rc^{[5]}_c= \{2,3,5\}, \ \Rc_s= \{3,4,6\}.
\end{align*}

 For this system, the tradeoff points $ (\text{sDoF}, \text{cDoF})=( 1 ,0)$, $ (\text{sDoF}, \text{cDoF})=( 0 ,2 )$  can be achieved by the  sensing-only  and communication-only schemes, respectively. We then propose a TIM ISAC scheme, achieving $ (\text{sDoF}, \text{cDoF})=(2/5, 2)$.
 
Let $t_0=10$. During these $10$ time slots,  the sensor 
 will obtain \(4\) effective observations through the sensing signals $\xv_1^{[0]}, \xv_2^{[0]}, \xv_3^{[0]}, \xv_4^{[0]},  \in \mathbb{C}^{3\times 1}$, satisfying \([\xv_1^{[0]}, \xv_2^{[0]}, \xv_3^{[0]}, \xv_4^{[0]}] [\xv_1^{[0]}, \xv_2^{[0]}, \xv_3^{[0]}, \xv_4^{[0]}]^H= \mathbf{I}_3\). 
In each period of $5$ time slots, we let each receiver recover $2$ communication message symbols, and let the sensor 
obtain $2$ effective observation. 
In the following, we illustrate our proposed scheme for the first  $5$ time slots, where each receiver $j\in \{1,\ldots,5\}$ should recover $2$ communication  messages 
$W_1^{[j]}, W_2^{[j]} \in \mathbb{C}$. 

Denote 
the transmission signals of  $j$-th transmitter by $\xv^{[j]} =[x^{[j]}(1),\ldots,x^{[j]}(5)]^T \in \mathbb{C}^{5 \times 1} $. 
Then 
for the transmitters  connected to the sensor, we design 
\begin{align*}
&\xv^{[3]} = \vv^{[3]} W_1^{[3]} + \vv^{[4]} W_2^{[3]} + \xv_s^{[3]},\\
&\xv^{[4]} = \vv^{[4]} W_1^{[4]} + \vv^{[5]} W_2^{[4]} + \xv_s^{[4]},\\
&\xv^{[6]} = \xv_s^{[6]};
\end{align*}
for the transmitters not connected to the sensor, we design 
\begin{align*}
&\xv^{[1]} = \vv^{[1]} W_1^{[1]} + \vv^{[2]} W_2^{[1]},\\
&\xv^{[2]} = \vv^{[2]} W_1^{[2]} + \vv^{[3]} W_2^{[2]},\\
&\xv^{[5]} = \vv^{[5]} W_1^{[5]} + \vv^{[6]} W_2^{[5]}.
\end{align*}
Note that $\vv^{[1]}, \ldots, \vv^{[6]} \in \mathbb{C}^{5 \times 1}$   represent $6$  random i.i.d. vectors, where any  $5$ of them   are linearly independent, and   $\xv_s^{[3]}, \xv_s^{[4]}, \xv_s^{[6]} \in \mathbb{C}^{5 \times 1}$  are the transmitted sensing signals to be determined later.

In this partially connected case, by exploiting the    network topology, we  align  the interference at  receivers $1$, $4$, and $5$ to a lower dimension. Let us consider  receiver $1$ as an example. The received signal after removing the dedicated sensing signals at time slot $t\in \{1,\ldots,5\}$ is
$
y^{[1]}(t) = \sum_{i \in \mathcal{R}_1=\{1,4,5\}}H^{[1i]} x^{[i]}(t) + z^{[1]}(t).
$ 
Hence, its received signals from the $5$ time slots can be formulated as:
\begin{align}
\yv^{[1]} &= 
\begin{bmatrix}
    y^{[1]}(1) \\
    \vdots \\
    y^{[1]}(5) \\
\end{bmatrix} 
= H^{[11]} (\vv^{[1]} W_1^{[1]} + \vv^{[2]} W_2^{[1]}) \nonumber \\ 
&\quad+ H^{[14]} (\vv^{[4]} W_1^{[4]} + \vv^{[5]} W_2^{[4]}) \nonumber \\
&\quad+ H^{[15]} \left(\vv^{[5]} W_1^{[5]} + \vv^{[6]} W_2^{[5]}\right) + \zv^{[1]} \nonumber \\
&= \underbrace{\vv^{[1]} H^{[11]} W_1^{[1]} + \vv^{[2]} H^{[11]} W_2^{[1]}}_{\text{desired signals}} \nonumber \\
&\quad + \vv^{[4]} H^{[14]} W_1^{[4]} + \vv^{[5]} \left(H^{[14]} W_2^{[4]} + H^{[15]} W_1^{[5]}\right) \nonumber \\
&\quad+ \vv^{[6]} H^{[15]} W_2^{[5]} + \zv^{[1]}    .    
\end{align}
The dimension of the designed signals is $2$, and the dimension of the interference is aligned to $ 3$. 
By the linear independence of $\vv^{[1]}, \vv^{[2]}, \vv^{[4]}, \vv^{[5]}, \vv^{[6]}$, receiver $1$ can recover $W_1^{[1]},W_2^{[1]} $ from the first  $5$ time slots. Similarly, each receiver can recover $2$ communication message symbols from the  first  $5$ time slots; thus the cDoF is $2$.

For the sensor, its received signal at time slot $t\in\{1,\ldots,5\}$ is
$
y^{[6]}(t) = \sum_{i \in \mathcal{R}_6=\{3,4,6\}}H^{[6i]} \xv^{[i]}(t) + z^{[6]}(t).
$
Hence, its received signals from the $5$ time slots can be formulated as:
\begin{align*}
&\yv^{[6]} = \begin{bmatrix}
y^{[6]}(1) \\
\vdots\\
y^{[6]}(5) \\
\end{bmatrix} 
= H^{[63]} \left(\vv^{[3]} W_1^{[3]} + \vv^{[4]} W_2^{[3]} + \xv_s^{[3]}\right) \nonumber \\
& + H^{[64]} \left(\vv^{[4]} W_1^{[4]} + \vv^{[5]} W_2^{[4]} + \xv_s^{[4]}\right)  + H^{[66]} \xv_s^{[6]} + \zv^{[6]} \nonumber \\
&= \vv^{[3]} H^{[63]} W_1^{[3]} + \vv^{[4]} \left(H^{[63]} W_2^{[3]} + H^{[64]} W_1^{[4]}\right) \nonumber \\
&  + \vv^{[5]} H^{[64]} W_2^{[4]}  + H^{[63]}\xv_s^{[3]} + H^{[64]}\xv_s^{[4]} + H^{[66]}\xv_s^{[6]} + \zv^{[6]}.  
\end{align*}

Note that $ [ \vv^{[2]} , \vv^{[3]} , \vv^{[6]} ] $ with dimension $5\times 3$ has $2$ linearly independent left null vectors with high probability;  
let $\vv_{0,1}, \vv_{0,2} \in \mathbb{C}^{1 \times 5}$ be the linearly independent  left null vectors. 
So
\begin{align*}
&\vv_{0,i}\yv^{[6]} \negmedspace=\negmedspace \vv_{0,i} \vv^{[3]} H^{[63]} W_1^{[3]} \negmedspace+\negmedspace \vv_{0,i} \vv^{[4]} \left(H^{[63]} W_2^{[3]} \negmedspace+ \negmedspace H^{[64]} W_1^{[4]}\right)  \\
&+ \vv_{0,i}\vv^{[5]} H^{[64]} W_2^{[4]}  + \vv_{0,i} H^{[63]}\xv_s^{[3]}    + \vv_{0,i} H^{[64]}\xv_s^{[4]} \\ &+
 \vv_{0,i} H^{[66]}\xv_s^{[6]}+ \vv_{0,i} \zv^{[6]}  \\ 
&=\begin{bmatrix} H^{[63]} & H^{[64]} & H^{[66]} \end{bmatrix} \begin{bmatrix} \vv_{0,i} \xv_s^{[3]} \\ \vv_{0,i} \xv_s^{[4]} \\ \vv_{0,i} \xv_s^{[6]} \end{bmatrix} + z'^{[6]},
\end{align*}
where $z'^{[6]}=\vv_{0,i}\zv^{[6]}, i \in \{1,2\}$.  
Hence,   the sensor can eliminate the interference from communication messages, and obtain  $2$ effective observations from the  first $5$ time slots. 

We then fix the values of $\xv_s^{[3]}, \xv_s^{[4]},\xv_s^{[6]}$. Recall that $x_1^{[0]},\dots,x_4^{[0]}$ has been selected before.  
Now we let $x_1^{[0]}=[ \vv_{0,1}\xv_s^{[3]} , \vv_{0,1}\xv_s^{[4]} , \vv_{0,1}\xv_s^{[6]} ], x_2^{[0]}=[ \vv_{0,2}\xv_s^{[3]} , \vv_{0,2}\xv_s^{[4]} , \vv_{0,2}\xv_s^{[6]} ]$; in other words, we solve $\xv_s^{[3]}$ with $5$ parameters satisfying $\vv_{0,i}\xv_s^{[3]}=x_{i,1}$, where $x_{i,1}$ is the first element of $\xv_i^{[0]}$, and similarly for $\xv_s^{[4]}$ and $\xv_s^{[6]}$.
Therefore, the sDoF  is $2/5$.

\end{example}

We next generalize Example~\ref{subsub:TIM noco 66} to achieve the tradeoff point  $(\text{sDoF}, \text{cDoF}) = \left(\frac{U+1}{\Ksf-D+U+1}, \frac{\Ksf(U+1)}{\Ksf-D+U+1}\right)$ in~\eqref{eq:noncooperative point}.
Each receiver or sensor is disconnected from the previous \( U \) and next \( D \geq U\) transmitters, and connected to other $\Ksf-U-D+1>0$ transmitters; thus the topology could be expressed as

\begin{align*}
& \mathcal{R}_c^{[k]} = \{ k, <\negmedspace k + D + 1 \negmedspace>_{\Ksf+1}, \ldots, <\negmedspace k - U - 1\negmedspace>_{\Ksf+1}\}, \\
& \Rc_s= \{\Ksf+1, <\negmedspace\negmedspace D+1\negmedspace\negmedspace>_{\Ksf+1}, <\negmedspace\negmedspace D+2\negmedspace\negmedspace>_{\Ksf+1},...,<\negmedspace\negmedspace \Ksf-U\negmedspace\negmedspace>_{\Ksf+1}\},
\end{align*}
 for each $ k \in\{1,\ldots,\Ksf\}$.

Let $t_0=(\Ksf-D-U+1)\left\lceil\frac{(\Ksf-D+U+1)}{U+1}\right\rceil$, during these $(\Ksf-D-U+1)\left\lceil\frac{(\Ksf-D+U+1)}{U+1}\right\rceil$ time slots, the sensor will obtain $N=(U+1)\left\lceil\frac{(\Ksf-D+U+1)}{U+1}\right\rceil$ effective observations, through the sensing signals $\xv_1^{[0]},\xv_2^{[0]},...,\xv_{N}^{[0]} \in \CC^{(\Ksf-D-U+1)\times1}$, satisfying  \([\xv_1^{[0]}, \xv_2^{[0]},..., \xv_{N}^{[0]}] [\xv_1^{[0]}, \xv_2^{[0]},..., \xv_N]^H= \textbf{I}_{(\Ksf-D+U+1)}\). 
In each period of $(\Ksf-D+U+1)$ time slots, we let each receiver decode $U+1$ communication message symbols, and let the sensor obtain $U$ effective observations. Next we will illustrate our proposed scheme for the first $\Ksf-D+U+1$ time slots, where each receiver $k \in \{1,2,...,\Ksf\}$ should recover $U+1$ communication message symbols $W_1^{[k]},\ldots,W_{U+1}^{[k]}\in \CC$. 

Denote  the transmitted signals of $j$-th transmitter,  $j\in \{1,\ldots,\Ksf+1\}$ by $\xv^{[j]} =[x^{[j]}(1),\ldots,x^{[j]}(\Ksf-D+U+1)]^T \in \mathbb{C}^{(\Ksf-D+U+1) \times 1} $.
For each  transmitter     connected to the sensor (i.e., $i\in \Rc_s \setminus \{\Ksf+1\}$), we design based on the cyclic coding,
\begin{align*}
    &\xv^{[i]} = \vv^{[i]} W_1^{[i]} + \vv^{[i+1]} W_2^{[i]} + \ldots + \vv^{[i+U]} W_{U+1}^{[i]} + \xv_s^{[i]}, \\
    & \xv^{[\Ksf+1]} = \xv_s^{[\Ksf+1]},
\end{align*}
 where  $\xv_s^{[j]}, j\in \mathcal{R}_s$ are the sensing signals to be determined later. For each
  transmitter  not connected to the sensor (i.e., $i\in \{1,\ldots,\Ksf+1\} \setminus \Rc_s$), we design based on the cyclic coding,
\begin{align*}
\xv^{[i]} = \vv^{[i]} W_1^{[i]} + \vv^{[i+1]} W_2^{[i]} + \ldots + \vv^{[i+U]} W_{U+1}^{[i]}.
\end{align*}
 Note that 
 $\vv^{[1]},...,\vv^{[U+\Ksf]}\in \CC^{\Ksf-D+U+1}$ represent $K+U$ randomly i.i.d vectors, where any $K-D+U+1$ of them are linearly independent. 
 
 For each $k\in \{1,\ldots,\Ksf\}$,  by removing the dedicated sensing signals from the received signals, the \(k\)-th  receiver obtains 

\begin{align*}
  &  \hat{\yv}^{[k]} = H^{[kk]}(\vv^{[k]} W_1^{[k]} +\negmedspace \cdots \negmedspace+ \negmedspace\vv^{[<k+U>_{\Ksf+1}]} W_{U+1}^{[k]}) \nonumber \\
     &+\negmedspace\negmedspace \sum_{i \in \Rc^{[k]}_c \setminus \{k\} } \negmedspace\negmedspace H^{[ki]}(\vv^{[i]} W_1^{[i]} \negmedspace+\negmedspace \ldots\negmedspace +\negmedspace \vv^{[<i+U>_{\Ksf+1}]} W_{U+1}^{[i]})\negmedspace +\negmedspace\zv^{[k]} \\
    &= \underbrace{H^{[kk]}(\vv^{[k]} W_1^{[k]}\negmedspace + \negmedspace\vv^{[k+1]} W_2^{[k]}\negmedspace +\negmedspace \ldots \negmedspace+\negmedspace \vv^{[<k+U>_{\Ksf+1}]} W_{U+1}^{[k]})}_{\text{desire signals}} \\
    &+ \underbrace{\sum^{k+\Ksf}_{i=k+D+1}\vv^{[i]}\sum^{\min(\Ksf -U-D,i)}_{s=\max(1,i-U)}H^{[ks]}W_{i-s+1}^{[s]}}_{\text{interference signals}} +\zv^{[k]}.
\end{align*}
Therefore, the dimension of the interference signals is aligned to \( K - D \). By the linear independence of the $\Ksf -D$ vectors $\vv^{[<k+D+1>_{\Ksf+1}]},   \vv^{[<k+D+2>_{\Ksf+1}]}, \ldots, \vv^{[<k+\Ksf>_{\Ksf+1}]}$, receiver $k$ can recover \( U+1 \) messages from the first \(\Ksf-D+U+1\) time slots. Thus, the cDoF is \(\frac{\Ksf(U+1)}{\Ksf-D+U+1}\).

For the sensor, the received signal is:

\begin{align*}
  &  \yv^{[\Ksf+1]} = H^{[\Ksf+1,\Ksf+1]} \xv_s^{[\Ksf+1]} + \sum^{\Ksf-U }_{i=D+1 }H^{[\Ksf+1,i]}\xv^{[i]} + \zv^{[\Ksf+1]} \nonumber \\
       &=H^{[\Ksf+1,\Ksf+1]} \xv_s^{[\Ksf+1]} \negmedspace + \negmedspace \sum^{\Ksf-D}_{i=1}\vv^{[i]}\sum^{\min(\Ksf-U-D,i)}_{s=\max(1,i-U)}H^{[\Ksf+1,s]}W_{i-s+1}^{[s]} \\
       &+\sum^{\Ksf-U}_{i=D+1}H^{[\Ksf+1,i]}\xv_s^{[i]} \negmedspace + \negmedspace \zv^{[\Ksf+1]}.
\end{align*}

The interference dimension from communication signals is the same as that for receivers, i.e., \(\Ksf-D\). Therefore, we can get $U$ linearly independent left null vectors $\vv_{0,1},\vv_{0,2},\dots,\vv_{0,U}$ of $\begin{bmatrix} \vv^{[D+1]} ,\vv^{[D+2]}, \ldots,  \vv^{[\Ksf]}\end{bmatrix}$ to remove the communication interference within \(\Ksf-D+U+1\) time slots. Thus

\begin{align*}
    \vv_{0,u} \yv^{[\Ksf+1]} &= H^{[\Ksf+1,\Ksf+1]} \vv_{0,u} \xv_s^{[\Ksf+1]}  \\ 
    \quad &+ \sum_{i=D+1}^{\Ksf-U} H^{[\Ksf+1,i]} \vv_{0,u} \xv_s^{[i]}  +  \vv_{0,u} \zv^{[\Ksf+1]} .
\end{align*}

We then fix the values of $\xv_s^{[\Ksf+1]},\xv_s^{[D+1]},\dots,\xv_s^{[\Ksf-U]}$, the orthogonal sensing signals $\xv_u^{[0]}, u\in \{1,2,\dots,U+1\}$ have been generated before. Let $\xv_u^{[0]}=\begin{bmatrix} \vv_{0,u}\xv_s^{[\Ksf+1]} , \vv_{0,u}\xv_s^{[D+1]} , \ldots , \vv_{0,u}\xv_s^{[\Ksf-U]} \end{bmatrix}^T$, we solve $\xv_s^{[\Ksf+1]}$ with $\Ksf-D+U+1$ elements satisfying $\vv_{0,u}\xv_s^{[\Ksf+1]}=x_{u,1}$, where $x_{u,1}$ is the first element of $\xv_u^{[0]}$, and similarly for $\xv_s^{[D+1]}, \dots, \xv_s^{[\Ksf-U]}$. $U+1$ effective observations of the channel can be acquired; all transmission can achieve an sDoF equal to $\frac{U+1}{\Ksf-D+U+1}$.

\subsection{Proof of Theorem~\ref{thm:KDco+1 region}: \( (\Ksf+1, d) \)  regular network}
\label{sub:regular network}
We then consider a bistatic ISAC system with heterogeneous connectivity, modeled as a $(\Ksf+1, d)$-regular network. For this network, we propose an ISAC scheme based on TIM achieving the tradeoff point  
 $(\text{sDoF}, \text{cDoF}) = \left( \frac{2}{d+1}, \frac{2\Ksf}{d+1} \right)$. Recall that the network topology could be expressed as
\begin{align*}
    & \Rc^{[k]}_c = \{k, <\negmedspace k+1 \negmedspace>_{\Ksf+1}, \ldots, <\negmedspace k+d-1 \negmedspace>_{\Ksf+1} \}, \\
    & \Rc_s = \{1,2,\ldots,d-1,\Ksf+1\},
\end{align*}
for each $k\in \{1,\ldots,\Ksf\}$.

Let $t_0 = (d+1)\left\lceil\frac{d}{2}\right\rceil$. In $(d+1)\left\lceil\frac{d}{2}\right\rceil$ time slots, the sensor will obtain $N = 2\left\lceil\frac{d}{2}\right\rceil$ effective observations, through the sensing signals $\xv_1^{[0]},\xv_2^{[0]},...,\xv_{N}^{[0]} \in \CC^{(d+1)\times1}$, satisfying  \([\xv_1^{[0]}, \xv_2^{[0]},..., \xv_{N}^{[0]}] [\xv_1^{[0]}, \xv_2^{[0]},..., \xv_{N}^{[0]}]^H= \textbf{I}_{(d+1)}\). 
In each period of $d+1$ time slots, we let each receiver decode $2$ communication message symbols, and let the sensor obtain $2$ effective observations. Next  we will  illustrate our proposed scheme for the first $d+1$ time slots, where each receiver $k \in \{1,2,...,\Ksf\}$ should recover $2$ communication message symbols $W_1^{[k]},W_{2}^{[k]}\in \CC$. 

Denote  the transmitted signals of $j$-th transmitter, $ j \in \{1,\ldots,\Ksf+1\}$ by $\xv^{[j]} =[x^{[j]}(1),\ldots,x^{[j]}(d+1)]^T \in \mathbb{C}^{(d+1) \times 1} $.
For the $d$ transmitters  connected to the sensor, let
\begin{align*}
    & \xv^{\left[j\right]} \negmedspace=\negmedspace \vv^{\left[j\right]} W_1^{\left[j\right]} \negmedspace+\negmedspace \vv^{\left[j+1\right]} W_2^{\left[j-d+1\right]} \negmedspace+\negmedspace \xv_s^{\left[j\right]}, \forall j\in \{1,2,\ldots,d-2\}, \\
    &\xv^{\left[d-1\right]} = \xv_s^{\left[d-1\right]} + \vv^{\left[d-1\right]} W_1^{\left[d-1\right]}, \\
    & \xv^{\left[\Ksf+1\right]} = \xv_s^{\left[\Ksf+1\right]} + \vv^{\left[1\right]} W_2^{\left[\Ksf-d+2\right]}.
\end{align*}
For the transmitters in $\{d, d+1, \ldots, \Ksf\}$, which are not connected to the sensor, let
\begin{align*}
    &\xv^{\left[j\right]} = \vv^{\left[j\right]} W_1^{\left[j\right]} + \vv^{\left[j+1\right]} W_2^{\left[j-d+1\right]}, \forall j\in \{d, d+1, \ldots, \Ksf\}. 
\end{align*}

For receiver $k\in \{d,\ldots,\Ksf-d+1\}$ which is only connected to the communication transmitter, the received signal is:

\begin{align*}
    \yv^{\left[k\right]} &= \sum_{j=k}^{k+d-1} H^{\left[kj\right]} \left( \vv^{\left[j\right]} W_1^{\left[j\right]} + \vv^{\left[j+1\right]} W_2^{\left[j-d+1\right]} \right)  + \zv^{[k]} \\
&=\underbrace{H^{\left[kk\right]} \vv^{\left[k\right]} W_1^{\left[k\right]} + H^{\left[k,k+d-1\right]} \vv^{\left[k+d\right]} W_2^{\left[k\right]}}_{{\rm desired \  signal}} \\
    \quad &+ \sum_{j=k+1}^{k+d-1} \vv^{\left[j\right]} \left( H^{\left[kj\right]} W_1^{\left[j\right]} + H^{\left[k,j-1\right]} W_2^{\left[j-d\right]} \right)+ \zv^{[k]}.
\end{align*}
The dimension of the desired signals  is 2, and the dimension of the interference signals  is $d-1$, we can achieve symmetric $\text{cDoF} = \frac{2}{d+1}$ for each receiver. Note that 
all interference at the  receivers originates from communication signals, while  sensing signals can be eliminated. The receiver that only receives communication signals experiences the strongest interference. For the receivers that receive both sensing and communication signals (the receivers not in $\{d,\ldots,\Ksf-d+1\}$), the interference dimension does not exceed \(d-1\); thus in $d+1$ time slots, each of them can also decode   $2$ communication message symbols, achieving the cDoF    \(\frac{2}{d+1}\) per receiver. 
Thus, the sum cDoF of the whole system is $\frac{2\Ksf}{d+1}$.

For the sensor, the received signal is:
\begin{align*}
    \yv^{[s]} &= \sum_{j=1}^{d-2} H^{[\Ksf+1,j]} \left( \vv^{[j]} W_1^{[j]} + \vv^{[j+1]} W_2^{[j-d+1]} + \xv_s^{[j]} \right) \\
    &\quad + H^{[\Ksf+1,\Ksf+1]} \left( \xv_s^{[\Ksf+1]} + \vv^{[1]} W_2^{[\Ksf-d+2]} \right) \\
    &\quad + H^{[\Ksf+1,d-1]} \left( \xv_s^{[d-1]} + \vv^{[d-1]} W_1^{[d-1]} \right) +\zv^{[\Ksf+1]} \\
    &=\sum^{d-1}_{j=1}\vv^{[j]}(H^{[\Ksf+1,j]}W_1^{[j]}+H^{[\Ksf+1,j-1]}W_2^{[j-d]})\\
    & \quad +\sum_{j=1}^{d} H^{[\Ksf+1,j-1]}\xv_s^{[j-1]}+\zv^{[\Ksf+1]}.
\end{align*}

Note that $[\vv^{\left[1\right]},  \vv^{\left[2\right]},  \ldots,  \vv^{\left[d-1\right]}]$ with dimension $(d+1)\times(d-1)$ has $2$ linearly independent left null vectors with high probability; let $\vv_{0,1}, \vv_{0,2} \in \CC^{1 \times (d+1)}$ be the linearly independent left null vectors. Thus 
\begin{align*}
    \vv_{0,i} \yv^{\left[s\right]} &=\sum_{j=1}^{d-1} H^{\left[K+1,j\right]}  \vv_{0,i} \xv_s^{\left[j\right]} +  H^{\left[K+1,K+1\right]} v_{0,i} \xv_s^{\left[K+1\right]} \\
    \quad &+ \vv_{0,i}\zv^{[\Ksf+1]} .
\end{align*}

We then fix the values of $\xv_s^{[\Ksf+1]},\xv_s^{[1]},\dots,\xv_s^{[d-1]}$, the orthogonal sensing signals $\xv_i^{[0]}, i =  \{1,2\}$ have been generated before. Let $\xv_i^{[0]}=\begin{bmatrix} \vv_{0,i}\xv_s^{[\Ksf+1]} , \vv_{0,i}\xv_s^{[1]} , \ldots , \vv_{0,i}\xv_s^{[d-1]} \end{bmatrix}^T$, we solve $\xv_s^{[\Ksf+1]}$ with $d+1$ elements satisfying $\vv_{0,i}\xv_s^{[\Ksf+1]}=x_{i,1}$, where $x_{i,1}$ is the first element of $\xv_i^{[0]}$, and similarly for $\xv_s^{[1]}, \dots, \xv_s^{[d-1]}$.   Consequently, $2$ effective observations of the channel can be acquired, all transmission can  achieve an sDoF equal to $\frac{2}{d+1}$.

\begin{remark}[Connection of the proposed ISAC schemes based on TIM]

In this section, we propose bistatic ISAC schemes based on TIM, which leverage stable network connectivity patterns to eliminate interference without requiring exact CSI. 
We consider two canonical network topologies: the $(\Ksf+1, d)$-regular network and the $(\Ksf+1, U, D)$ neighboring antidotes network, where the specific topology is determined by the underlying network structure.
\begin{itemize}
    \item These two networks give rise to two respective schemes. Although the connectivity patterns differ, both networks exhibit cyclic topological structures. Therefore, we adopt a cyclic coding strategy in both cases to exploit the network connectivity and align the interference signals at the receivers, thereby reducing the effective interference dimension.
    \item However, the detailed construction of the linear coding schemes (based on subspace alignment) depends on the specific topology. In the $(\Ksf+1, d)$-regular network, transmitters are cooperative, and each receiver is connected to a sequence of $d$ consecutive transmitters in the topology. This allows each transmitter to send one symbol for its own target receiver and another for the receiver with the maximal topological distance, facilitating interference separation at the receivers.

In contrast, in the $(\Ksf+1, U, D)$ neighboring antidotes network, each receiver is disconnected from a number of nearby interfering transmitters in the topology. This disconnection enables a cyclic coding design such that the signals from interfering transmitters are aligned into a lower-dimensional subspace at each receiver.
\end{itemize}
\end{remark}

\section{Simulation Results}
\begin{figure*}
    \centering
    \includegraphics[width=0.8\linewidth]{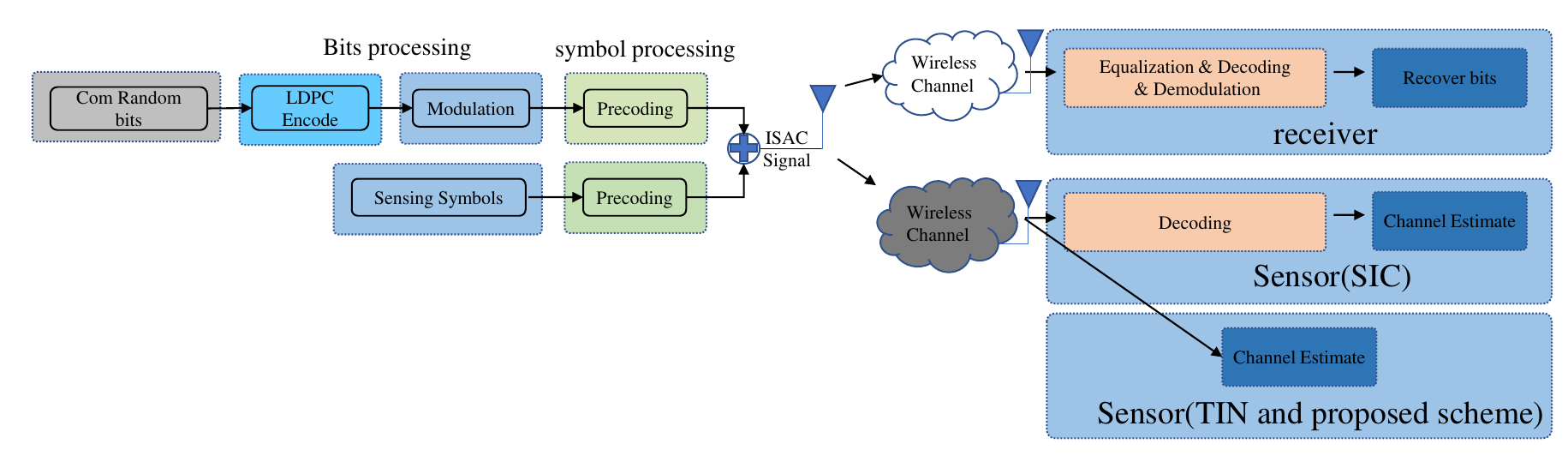}
    \caption{\small Simulation flowchart.}
    \label{fig:simuprocess2}
\end{figure*}

\begin{table}
\centering
 
\caption{Simulation Parameters}
\label{tab:sim_params}
\begin{tabular}{llc}
\toprule \textbf{Parameter} & \textbf{Value} \\
\midrule
SNR range (dB) & $[-5,\ 35]$ (step 5 dB) \\
 Number of transmitted symbols & 1000 \\
 Transmit power (dBm) & $30$ dBm \\
  Path loss exponent & 3.5 \\
 Number of Monte Carlo simulations & $100$ \\
 Carrier frequency & $2.4$ GHz \\
 Sampling frequency & $20$ MHz \\
 Symbol rate & $1$ MHz \\
 LDPC code rate & $0.5$ \\
 Samples per symbol & $20$ \\
 Modulation order & $4$ (DQPSK) \\
 Rolloff factor & $0.25$ \\
 Filter span (symbols) & $6$ \\
 Comm. user distances & Uniformly $[50,100]$ m \\
 Sensing target distances & Uniformly $[10,50]$ m \\
\bottomrule
\end{tabular}
\end{table}

We simulate proposed schemes for the  bistatic ISAC systems with heterogeneous coherent times and
the interference channel   in Example~\ref{subsub:BIA IC case}, and  with   heterogeneous connectivity and the interference channel in Example~\ref{subsub:TIM noco 66}. 
 The simulation flowchart is illustrated in Fig.~\ref{fig:simuprocess2} and the simulation parameters are provided in Table~\ref{tab:sim_params}. All the source codes can be found in \cite{liu2025ISAC}.

{\bf Overview of the simulation.} During the simulation, the signal-to-noise ratio (SNR) $P/\sigma^2$ is varied from $[-5,\ 35]$~dB in increments of $5$~dB. $P$ represents the transmission power;  $\sigma^2$ represents the noise power at the receiver (and also the noise power at the sensor). 
communication receivers are uniformly distributed within a range of $[50,\ 100]$~meters, while sensing targets are placed within $[10,\ 50]$~meters. Large-scale path loss follows a distance-based model with a path loss exponent of 3.5, and small-scale fading is modeled using a Rayleigh distribution,  where the complex channel coefficient \( h = h_R + j h_I \) has independent real and imaginary parts \( h_R, h_I \sim \mathcal{N}(0, 1/2) \). As a result, the envelope \( |h| \) follows a Rayleigh distribution.   For the sensing task, we use the LS estimation method to estimate the channel, given by $\hat{\mathbf{H}} = \mathbf{Y} \mathbf{X}^H (\mathbf{X} \mathbf{X}^H)^{-1}$,
where \(\hat{\Hm}\) represents the estimated channel, \(\Xm\) represents the sensing signal, and \(\Ym\) represents the received signal. The CEE is defined as $\text{CEE} = \| \Hm - \hat{\Hm} \|^2$, where \(\Hm\) is the actual channel state, $\hat{\Hm}$ is the estimated channel state, $\|\cdot\|$ denotes the norm of the vector, commonly the Euclidean norm.
In the heterogeneous coherence times scenario, the transmitter is assumed to have a statistic prior of sensing channels. In contrast, for the heterogeneous connectivity scenario, the transmitter only knows the network topology and similarly lacks sensing channel information. The dedicated sensing waveform is deterministic and pre-known at both the transmitter and the receiver sides. Each Monte Carlo realization transmits 1000 symbols, and the final performance metrics are averaged over 100 independent trials to ensure statistical robustness. 
 
{\bf Generation of the dedicated sensing signal.} 
We set the power of the dedicated sensing signal $P_s=P/2$ as half of the total power. 
As shown in Footnote~\ref{foot:orthogonal}, since  the statistic CSI is known at the $\msf'$ single antenna transmitters connected to the sensor,  the transmitted deterministic sensing signal can be $\begin{bmatrix}
\xv_1^{[0]}, \xv_2^{[0]}, \ldots, \xv_N^{[0]}
\end{bmatrix} = \Wm\Sm_D$ with a deterministic training signal $\Sm_D\in \mathbb{C}^{\msf' \times N}$ satisfying 
$\Sm_D \Sm_D^H = \mathbf{I}_{\msf'}$ and a precoding matrix $\Wm \in \mathbb{C}^{\msf' \times \msf'}$ to   minimize detection/estimation errors.   \(\mathbf{W}\) is obtained by solving the following optimization problem based on the Linear Minimum Mean Square Error (LMMSE) criterion~\cite{lu2024random}, where the goal is to minimize the mean squared error (MSE) subject to a transmit power constraint:
\begin{equation}
    \min_{\mathbf{W} \in \mathcal{A}} J_{\text{LMMSE}} = \text{tr} \left[ \left( \mathbf{R}_H^{-1} + \frac{1}{\sigma^2 } \mathbf{W} \mathbf{W}^H \right)^{-1} \right],
\end{equation}
where \(\mathbf{R}_H\) denotes the statistical channel correlation matrix. By performing eigendecomposition of \(\mathbf{R}_H = \mathbf{Q}\mathbf{\Lambda}\mathbf{Q}^H\) and applying the water-filling principle, the optimal solution can be expressed as:
\begin{equation}
    \mathbf{W} = \sqrt{\frac{\sigma^2 }{N}} \, \mathbf{Q} \left[ \left( \mu_0 \mathbf{I}_{\msf'} - \mathbf{\Lambda}^{-1} \right)^+ \right]^{\frac{1}{2}},
\end{equation}
where \(\mathbf{\Lambda}^{-1} = \text{diag}(\lambda_1^{-1}, \lambda_2^{-1}, \ldots, \lambda_{m'}^{-1})\), \((\cdot)^+\) denotes the positive part function \((\mu_0 - \lambda_i^{-1})^+ = \max(\mu_0 - \lambda_i^{-1}, 0)\), and the water level \(\mu_0\) is chosen to satisfy the power constraint:
\begin{equation}
    \frac{\sigma^2 }{N} \sum_{i=1}^{\msf'} (\mu_0 - \lambda_i^{-1})^+ = P_s.
\end{equation}

{\bf Generation of the dedicated communication signal.} 
Once the sensing signal is precoded, it is superimposed with the modulated communication signal in the time domain. 
We set the power of the dedicated communication signal $P_c=P/2$, also as   half of the total power. 
The resulting baseband signal is oversampled at a rate of 20 samples per symbol. The system employs Low-Density Parity-Check (LDPC) code channel coding followed by DQPSK modulation to encode the communication bits. The  LDPC code rate is $0.5$, and the symbol rate is set to $1$~MHz, corresponding to a sampling frequency of $20$~MHz. Pulse shaping is applied using a raised cosine filter with a roll-off factor of $0.25$ and a span of $6$ symbols. The carrier frequency is fixed at $2.4$~GHz. 

 {\bf Benchmark  schemes.} For the sake of comparison, we consider two benchmark  schemes that differ in how the sensor handles the communication signal, where the TIN scheme treats the communication signal as noise without decoding, and the SIC scheme fully decodes the communication messages and then removes them to improve sensing performance. At the transmitter side, both benchmark schemes directly send the sum of the dedicated sensing   and communication signals, while the proposed schemes additionally precode the   dedicated sensing signal and communication signals based on BIA or TIM.

For   both of benchmark schemes, the communication receiver directly removes the interference of the dedicated sensing signal and then decodes the message. 
Let us then introduce the main steps of the benchmark schemes at the sensor side.
The TIN benchmark scheme  
 consists of the following steps:
\begin{enumerate}
    \item Matched filtering and sampling: The received ISAC waveform undergoes downconversion, matched filtering (using a square-root raised cosine filter), and symbol-rate sampling to extract the complex baseband symbols. In this method, interference from communication symbols is not mitigated but instead treated as noise.
    \item Channel estimation: Using estimate method to estimate the channel.
\end{enumerate}
The SIC benchmark scheme 
consists of the following steps: 
\begin{enumerate}
    \item Matched filtering and sampling: The received ISAC signal is downconverted and passed through a matched filter (e.g., square-root raised cosine). It is then sampled at the symbol rate to extract complex-valued baseband symbols.
    \item Differential phase detection: For each extracted symbol \( y_n \), the phase difference with its predecessor is computed as
    \[
    \Delta\phi_n = \angle(y_n) - \angle(y_{n-1}),
    \]
    followed by normalization into the range \( (-\pi, \pi] \) to account for phase wrapping.
    \item Nearest constellation matching: The normalized phase difference \( \Delta\phi_n \) is compared against the ideal DQPSK phase shifts \( \{0, \pi/2, \pi, 3\pi/2\} \). The detected symbol index is determined by selecting the closest phase:
    \[
    \hat{a}_n = \arg\min_{k \in \{0,1,2,3\}} \left| \Delta\phi_n - \theta_k \right| \mod 2\pi,
    \]
    where \( \theta_k = \frac{2\pi k}{M} \), and \( M = 4 \) for DQPSK.
    \item Symbol remapping: The detected indices \( \hat{a}_n \in \{0,1,2,3\} \) are mapped to complex DQPSK constellation points .
    \item ISAC signal reconstruction and channel estimation: The recovered communication symbols and sensing signals are used to reconstruct the ISAC signal, enabling subsequent radar channel estimation.
\end{enumerate}

\begin{figure}[htbp]
    \centering
    \includegraphics[width=0.4\textwidth]{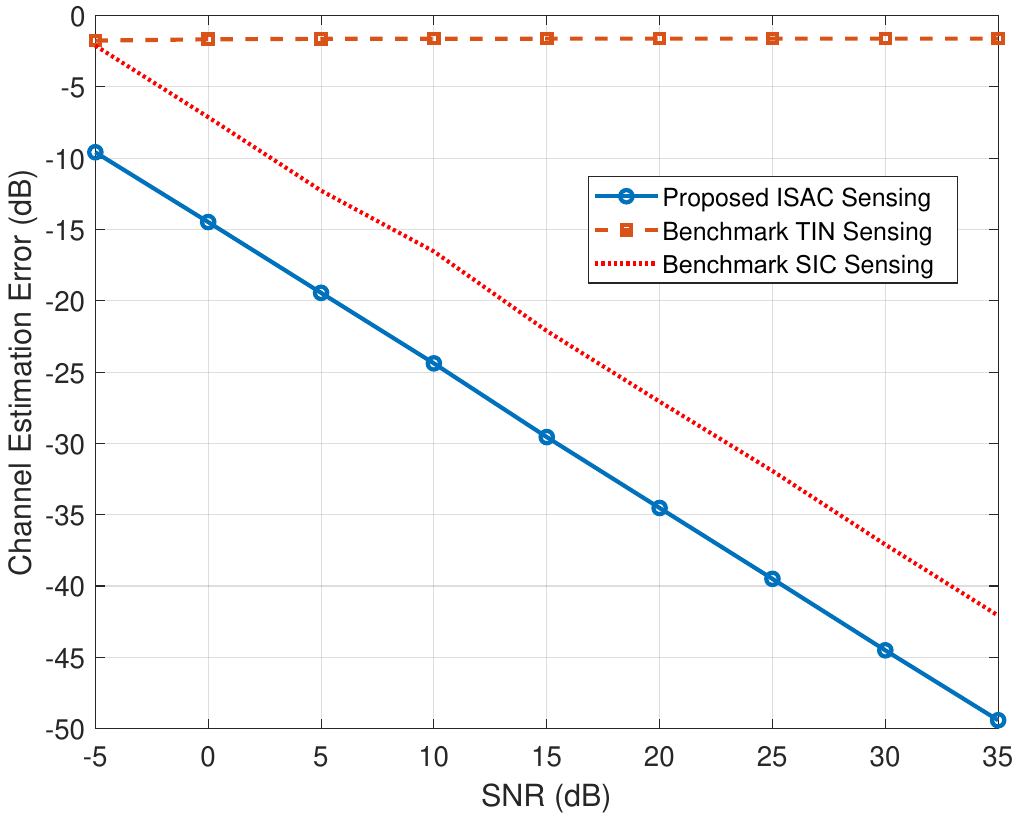}
    \caption{\small Simulation result of BIA.}
    \label{fig:simuBIA}
\end{figure}

\begin{figure}[htbp]
    \centering
    \includegraphics[width=0.4\textwidth]{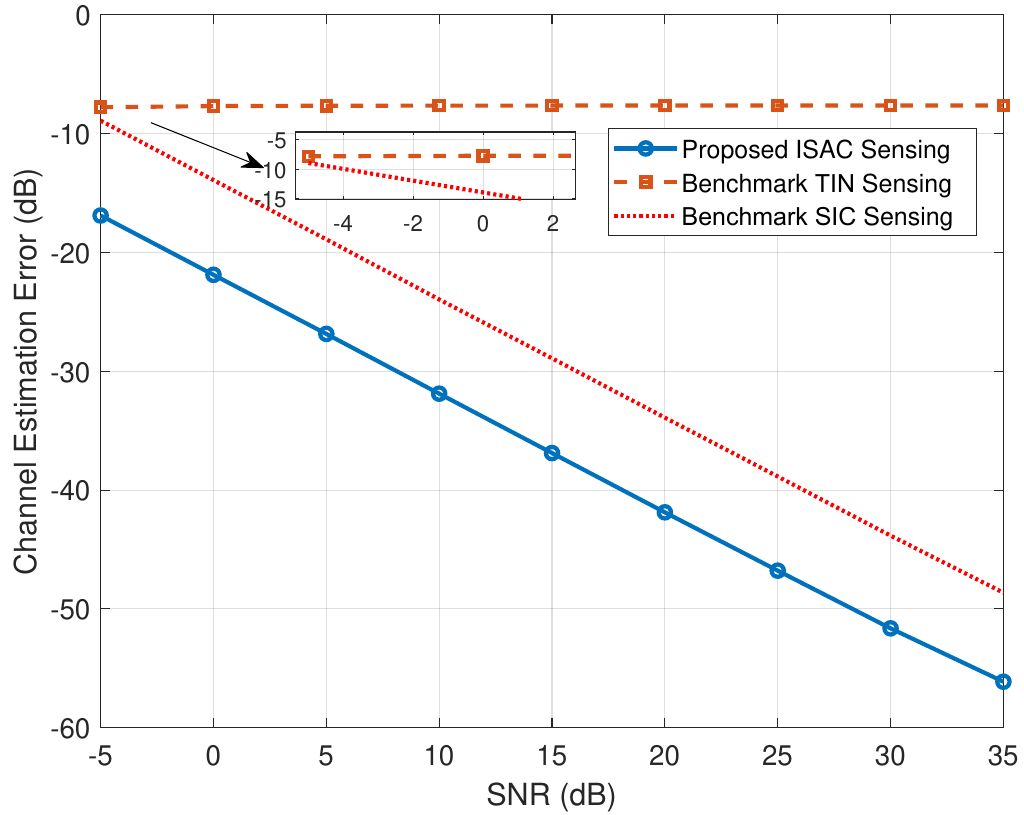}
    \caption{\small Simulation result of TIM.}
    \label{fig:simuTIM}
\end{figure}

\subsection{Heterogeneous Coherence Times}

 For the  bistatic ISAC systems with heterogeneous coherent times and
the interference channel   in Example~\ref{subsub:BIA IC case}, 
 as shown in Fig.~\ref{fig:simuBIA}, the proposed method demonstrates superior robustness to interference, especially in the high-SNR regime. It significantly reduces the CEE  as SNR increases. In contrast, the TIN benchmark scheme   treats the communication signal as noise, which dominates sensing performance at high power levels, resulting in persistently high estimation error even as noise decreases.

Quantitatively, our method achieves an estimation accuracy gain of $[7.83, 47.81]$~dB over the TIN benchmark scheme. Compared with the SIC benchmark scheme which performs noncoherent detection followed by signal subtraction—the proposed scheme also shows consistent improvement. While the SIC benchmark scheme suffers from poor detection at low SNR and improves at higher SNR, our method still outperforms it by $[7.17, 7.85]$~dB in the high-SNR regime.

The proposed scheme exhibits an overall complexity of $O(\Ksf^3)$, primarily due to the LS-based channel estimation, with an additional differencing step of only $O(\Ksf)$ complexity. The SIC benchmark scheme also incurs $O(\Ksf^3)$ complexity, comprising $O(\Ksf^2)$ for noncoherent detection and $O(\Ksf^3)$ for channel estimation, along with added overhead from signal reconstruction. The TIN benchmark scheme skips symbol decoding and directly performs channel estimation, resulting in a simpler structure but the same asymptotic complexity of $O(\Ksf^3)$. Thus, while all methods share the same complexity order, their practical implementation overheads differ.

It is important to highlight that all three schemes yield comparable communication performance (e.g., in terms of rate or BER). In other words, with the same communication performance, our proposed method significantly improves the sensing performance.


\subsection{Heterogeneous Connectivity}
For the  bistatic ISAC systems  with   heterogeneous connectivity and the interference channel in Example~\ref{subsub:TIM noco 66}, as   shown in Fig.~\ref{fig:simuTIM},  
compared to the TIN benchmark scheme, our proposed approach achieves a significant improvement in channel estimation accuracy, with a gain ranging from $[7.49,48.19]$~dB.  
In addition, compared to the SIC benchmark scheme, our approach still exhibits a performance advantage of $[7.65,7.99]$~dB under high SNR conditions.
As in  the scenario of heterogeneous coherence times, the SIC benchmark scheme performs poorly in low SNR regimes due to residual interference and limited noise suppression capability, and as the SNR increases, it  gradually improves and becomes more effective in mitigating communication-induced interference, thus reducing the channel estimation error.

 The proposed scheme incurs a total complexity of $O\left((\Ksf + U) \cdot L^2 + L^3\right)$ with $L=\Ksf-D+U+1$, where the first term arises from precoding construction and the second from sensing signal design and LS-based channel estimation. The SIC benchmark scheme involves noncoherent detection and LS estimation, with overall complexity $O(L^3)$, while the TIN benchmark scheme bypasses symbol detection and relies solely on LS estimation, also at $O(L^3)$. Although all methods share the same asymptotic complexity order with respect to $L$, the proposed scheme requires additional steps for structured signal design, whereas the TIN benchmark scheme offers the lowest implementation overhead.
 
\begin{remark}[Discussion on the theoretic assumptions]
\label{rem:discussion}
The theoretical results in Section~\ref{sec:main results} are established under the following common assumptions: (1) the transmitters have perfect CSI or knowledge of the network topology of the communication channel, and the communication receivers have perfect CSIR; (2) 
the communication and sensing channels have heterogeneity, either in coherence time or in network topology; (3) the tradeoff between sDof and cDof is considered.


However, in our  simulations, we discard assumption (3) by considering the practical sensing and communication metrics. It can be seen from the simulation results that, our interference elimination strategies based  on BIA and TIM can also improve the ISAC  performance on practical  metrics (e.g., estimation error, communication rate/error).


Fundamentally, all our schemes exploit the heterogeneity inherent in ISAC systems—such as differences in coherence time and network connectivity structures—to enable effective interference elimination and performance enhancement. In the ongoing work, we are extending this interference elimination strategy to more general ISAC systems,  under more relaxed assumptions.

\end{remark}

\section{Conclusion}
This paper explored the interference management issue in bistatic ISAC systems, specifically how to 
 effectively cancel the interference from the communication signals to the sensor, by    applying   
 the BIA and TIM strategies. The key is to leverage   the heterogeneity (on the coherence times or on the connectivity) between the sensor and receivers.
  By using the communication and  sensing degrees  of freedom as the metric,  the proposed schemes provide improved 
    tradeoff   points, compared to the time-sharing between the communication-only and sensing-only points. 
  Simulations on the proposed schemes and on the existing TIN, SIC schemes are provided, showing that the proposed schemes  reduce the sensing CEE, while maintaining the same communication performance.

\appendices

\section{Proof of Theorem~\ref{thm:MUMIMO}: MU-MIMO}
\label{pro:MUMIMO}
Next, we  consider the bistatic ISAC systems with heterogeneous coherence times and  the MU-MIMO channel including one transmitter with $\msf$  antennas, $\Ksf$ multi-antenna  receivers (receiver $k=1,\ldots,\Ksf$ with $\nsf_k$ antennas), and one single-antenna sensor.  By Theorem~\ref{thm:MUMIMO}, the proposed BIA scheme achieves the tradeoff point
$(\text{sDoF}, \text{cDoF}) = \left(\frac
{\left\lceil\frac{\msf}{\sum_{k=1}^{\Ksf}\nsf_k}\right\rceil-1}
{\left\lceil\frac{\msf}{\sum_{k=1}^{\Ksf}\nsf_k}\right\rceil}, 
\frac{\msf} {\left\lceil\frac{\msf}{\sum_{k=1}^{\Ksf}\nsf_k}\right\rceil} \right)$.
Note that when $\msf \leq  \sum_{k=1}^{\Ksf}\nsf_k$, the tradeoff point becomes $(0, \msf)$, which could be simply achieved by the communication-only scheme. Hence, we only consider the case $\msf > \sum_{k=1}^{\Ksf}\nsf_k$. 
For ease of notation, we define that $\left\lceil\frac{\msf}{\sum_{k=1}^{\Ksf}\nsf_k}\right\rceil := \bsf$,  and $<\msf>_{\sum_{k=1}^{\Ksf}\nsf_k} := \rsf$. The number of receivers, denoted by \( \Ssf \) (which is determined by $\sum_{i=1}^S{\nsf_i} \leq \rsf  <\sum_{j=1}^{S+1}{\nsf_i} $), means that can perfectly exploit the multi-antenna diversity gain at the receiver during each transmission time slot (perfect meaning that the number of symbols decodable at each moment is equal to the number of antennas), $\qsf=\rsf-\sum^{\Ssf}_{i=1}\nsf_i$. 

Let $t_0= (\bsf\left\lceil\frac{\msf}{\bsf-1}\right\rceil )$, and let the receiver $k\ \in\ \{ 1,2,...,\Ksf\}$ decode messages $W_{j,i,t}^{[k]},i \in \{1,2,...,\nsf_k\}$ at time slot $t\ \in\ \{1,2,...,\bsf-1\}$ in the $j$-th transmission, receiver $k\ \in\ \{ 1,2,...,\Ssf\}$ decode messages $W_{j,i,\bsf}^{[k]}, i \in \{1,2,...,\nsf_k\}$ at time slot $\bsf$ in the $j$-th transmission in addition, and receiver $\Ssf+1$ decode messages $W_{j,i,\bsf}^{[\Ssf+1]}, i \in \{1,2,...,\qsf\}$ at time slot $\bsf$ in the $j$-th transmission in addition. Communication signals in the $\bsf$ time slots by  transmitter be \(\textbf{x}_c(1)=\textbf{x}_c(2)=\cdots=\textbf{x}_c(\bsf)=\xv_c \in \CC^{m \times 1}\) (it is the same as previous section), $\bsf$ time slots represent the duration of one transmission in the system. We accumulate valid observation data for the sensing task during the $\left\lceil\frac{\msf}{\bsf-1}\right\rceil $transmissions. For simplicity and without loss of generality, we focus on a specific transmission process. For each time slot $t$, we design the communication signals by transmitter with $m$ anttenas as :

\begin{align*}
    \xv_c(1)=\ldots=\xv_c(b) &= \sum^{\bsf-1}_{t=1}\sum^{\nsf_k}_{i=1}\sum^{\Ksf}_{k=1}\vv_{i,t}^{[k]}W_{i,t}^{[k]} \\
    \quad&+ \sum^{\nsf_k}_{i=1}\sum^{\Ssf}_{k=1}\vv_{i,\bsf}^{[k]}W_{i,\bsf}^{[k]}   
     +\sum^{\qsf}_{i=1}\vv_{i,\bsf}^{[\Ssf+1]}W_{i,\bsf}^{[\Ssf+1]} 
\end{align*}
where $\vv_{i,t}^{[k]}:k\in \{1,\ldots,\Ksf\},i\in \{1,\ldots,\nsf_k\},t \in \{1,\ldots,\bsf-1\}$,$\vv_{i,\bsf}^{[k]}:k\in \{1,\ldots,\Ssf\},i\in \{1,\ldots,\nsf_k\}$,$\vv_{i,\bsf}^{[\Ssf+1]}:i\in \{1,\ldots,\qsf\}$ are the precoding vectors to be determined later, each with dimension $\msf \times 1$. At time slots $t \in\{1,\ldots,\bsf\}$, the signal received by the $k \in \{1,\ldots,\Ksf\}$-th receiver is:
\begin{equation*}
    \yv^{[k]}=\Hm^{[k1]}(t)(\xv_c(t)+\xv_s(t)) + \zv^{[k]}(t) ,
\end{equation*}
we can remove the sensing signal from the received signal, resulting in an estimate siganl that only concerns the communictaion signals:
\begin{equation*}
    \hat{\yv^{[k]}} = \begin{bmatrix}
    h^{[11]}(t) & h^{[12]}(t) & \cdots & h^{[1\msf]}(t) \\
    h^{[21]}(t) & h^{[22]}(t) & \cdots & h^{[2\msf]}(t) \\
    \vdots & \vdots & \ddots & \vdots \\
    h^{[\nsf_k 1]}(t) & h^{[\nsf_k 2]}(t) & \cdots & h^{[\nsf_k \msf]}(t)
\end{bmatrix}\xv_c(t) + \zv^{[k]}(t) ,
\end{equation*}
where $h^{[ji]}(t)\in \ \CC$ denotes the channel coefficient from the $i$-th antenna of the transmitter to the $j$-th antenna of user $k$ at time slot $t$ . The precoding vectors are designed to ensure that 
\begin{itemize}
    \item for each $k\in \{1,\ldots,\Ssf\}$ , $i\in \{1,\ldots,\nsf_k\}$and $t\in \{1,\ldots,\bsf\}$, the $k$-th receiver can decode $W^{[k]}_{i,t},i\in \{1,\ldots,\nsf_k\}$ from $\hat{y}^{[k]}(t)$;
    \item for the $\Ssf+1$-th receiver can decode $W^{[\Ssf+1]}_{i,t},i\in \{1,\ldots,\nsf_{\Ssf+1}\},t\in \{1,\ldots,\bsf-1\} $ and $W^{[\Ssf+1]}_{i,\bsf},i\in \{1,\ldots,\qsf\}$ from $\hat{y}^{[\Ssf+1]}(t)$;
    \item for each $k\in \{\Ssf+2,\ldots,\Ksf\}$ , $i\in \{1,\ldots,\nsf_k\}$and $t\in \{1,\ldots,\bsf-1\}$, the $k$-th receiver can decode $W^{[k]}_{i,t},i\in \{1,\ldots,\nsf_k\}$ from $\hat{y}^{[k]}(t)$;
\end{itemize}

To satisfy the above conditions, we propose the following design on the precoding vectors:
Concatenate the channel matrices of all users at the time slots when they can correctly decode, column by column, we can obtain:
    $$
    \Hm_{all} =
\begin{bmatrix}
    \Hm^{[11]}(1) \\ \vdots \\ \Hm^{[11]}(\bsf) \\ 
    \vdots \\ \Hm^{[\Ssf1]}(\bsf)\\
    \Hm^{[\Ssf+1,1]}(1) \\ \vdots \\ \Hm^{[\Ssf+1,1]}(\bsf)[1:\qsf,:] \\
    \Hm^{[\Ssf+2,1]}(1) \\ \vdots \\ \Hm^{[\Ssf+2,1]}(\bsf-1) 
    \\ \vdots \\ \Hm^{[\Ksf,1]}(\bsf-1)    
\end{bmatrix}\in \CC^{\msf \times 1},
$$
where $\Hm^{[\Ssf+1,1]}(\bsf)[1:\qsf,:]$ represents taking only the first $\qsf$ rows of $\Hm^{[\Ssf+1,1]}(\bsf)$, which means that in time slot $\bsf$, we are utilizing only the first $\qsf$ antennas of the \( \Ssf+1 \) communication receivers.

\begin{itemize}
    \item Design on $\vv_{i,t}^{[k]}$ for $k\in \{1,\ldots,\Ssf\},i\in \{1,\ldots,\nsf_k\},t \in \{1,\ldots,\bsf\}$. We let $\vv_{i,t}^{[k]},i \in \{1,\ldots,\nsf_k\}$ be a right null vector of the matrix $\Hm_{all} \backslash \Hm^{[k1]}(t) $ with dimension $(\msf-\nsf_k) \times \msf$, there are $\nsf_k$ right null vertors in its null space, which is the precoding vector $\vv_{i,t}^{[k]},i \in \{1,\ldots,\nsf_k\}$.

    \item Design on $\vv_{i,t}^{[k]}$ for $k\in \{\Ssf+1,\ldots,\Ksf\},i\in \{1,\ldots,\nsf_k\},t \in \{1,\ldots,\bsf-1\}$. We let $\vv_{i,t}^{[k]},i \in \{1,\ldots,\nsf_k\}$ be a right null vector of the matrix $\Hm_{all} \backslash \Hm^{[k1]}(t) $ with dimension $(\msf-\nsf_k) \times \msf$, there are $\nsf_k$ right null vertors in its null space, which is the precoding vector $\vv_{i,t}^{[k]},i \in \{1,\ldots,\nsf_k\}$.
    
    \item Design on $\vv_{i,\bsf}^{[\Ssf+1]}$ for $i\in \{1,\ldots,\qsf \}$ for the $\Ssf+1$-th receiver at time slot $\bsf$. We let $\vv_{i,\bsf}^{[\Ssf+1]}$ be a right null vector of the matrix $\Hm_{all} \backslash \Hm^{[\Ssf+1,1]}(\bsf)[1:\qsf,:] $ with dimension $\msf-\qsf$, there are $\qsf$ right null vertors in its null space, which is the precoding vector $\vv_{i,\bsf}^{[\Ssf+1]}$.
\end{itemize}

By the above selection on the precoding vectors, one can check that the decodability conditions could be satisfied with high probability. For example, let us focus on receiver $1$. Its received signal after removing the sensing signal at time slot $t\in \{1,\ldots,\bsf\}$ is  
\begin{align}
    &\hat{y}^{[1]}(t) = \Hm^{[11]}(t)\xv_c(t) +z^{[1]}(t) \nonumber\\
    &=\Hm^{[11]}(t)(\sum^{\bsf-1}_{t=1}\sum^{\nsf_k}_{i=1}\sum^{\Ksf}_{k=1}\vv_{i,t}^{[k]}W_{i,t}^{[k]} \nonumber \\
    \quad&+ \sum^{\nsf_k}_{i=1}\sum^{\Ssf}_{k=1}\vv_{i,\bsf}^{[k]}W_{i,\bsf}^{[k]}   
     +\sum^{\qsf}_{i=1}\vv_{i,\bsf}^{[\Ssf+1]}W_{i,\bsf}^{[\Ssf+1]} )+z^{[1]}(t). \label{eq:MIMOy1}
\end{align}

It can be checked that the product of $\Hm^{[11]}(t)$ and each precoding vector in~\eqref{eq:MIMOy1} (except $\vv^{[1]}_{i,t},i \in \{1,\ldots,\nsf_1\}$) is $0$. Hence, receiver~$1$ can recover $W^{[1]}_{i,t},i \in \{1,\ldots,\nsf_1\}$ with high probability. Thus the $\text{cDoF} = \frac{\sum^{\Ssf}_{i=1}\bsf n_i+\sum^{\Ksf}_{j=\Ssf+1}(\bsf-1)n_j +\qsf}{\bsf}=\frac{m}{\bsf}$

We then  design the transmitted dedicated sensing signals $\xv_s(t)$ for $t\in \{1,\ldots,\bsf\}$  as follows (recall that $\xv_1^{[0]}, \ldots, \xv_{\bsf}^{[0]}$ have been selected before)
\begin{align*}
&\xv_s(1)=\sum^{\bsf-1}_{i=1} \xv_i^{[0]}, \ \  \xv_s(t)=\sum^{\bsf-1}_{i=1,i \neq t-1} \xv_i^{[0]}, \ \forall t\in \{2,\ldots,\bsf\}.
\end{align*} 
At each time slot $t\in \{1,\ldots,\bsf\}$, the sensor receives 
\begin{equation*}
    y_s^{[\Ksf+1]}(t) = \hv^{[\Ksf+1]}(t) (\xv_c(t)+\xv_s(t)) +z_s(t),
\end{equation*}
where \(\hv^{[\Ksf+1]}(t) = [ h^{[\Ksf+1,1]}(t) , \dots , h^{[\Ksf+1,\msf]}(t) ]\), and $h^{[\Ksf+1,i]}(t)$ represents  the channel coefficient from $i$-th antenna of the transmitter to the sensor. 

Since $\hv^{[\Ksf+1]}(1)=\cdots=\hv^{[\Ksf+1]}(\bsf)$ and $\xv_c(1)=\cdots=\xv_c(\bsf)$,
 the sensor can subtract the received signal at time slot \(t\in \{2,\ldots,\bsf\}\) from the received signal at the first time slot:
\begin{align*}
    \yv_s^{[\Ksf+1]}(1) - \yv_s^{[\Ksf+1]}(t) 
    = \hv^{[\Ksf+1]}(1)\xv_{t-1}^{[0]}+z_s(1)-z_s(t),
\end{align*}
thereby canceling the interference caused by the communication signals and obtaining an effective observation of \(\xv_{t-1}^{[0]}\). Over \(\bsf\) time slots, a total of \(\bsf-1\) effective observations can be made, and the sensing degrees of freedom (SDoF) is \((\bsf-1)/\bsf\).

By the above selection  we can select vectors from its null space to code the desired signals $W_{t,j}^{[k]}$. Through this zero-forcing precoding, we can achieve the placement of interference signals in the null space of the channel at the receiver, thereby enabling decoding.
As a result, over $\bsf$ time slots, $\msf$ symblos are totally decoded; thus the achieved communication DoF is $\frac{\msf}{\bsf}=\frac{\msf}{\left\lceil\frac{\msf}{\sum_{k=1}^{\Ksf}\nsf_k}\right\rceil}$.
At the sensor, since the communication signal remain constant across a time slots, over $\bsf\left\lceil\frac{\msf}{\bsf}\right\rceil$
time slots, the sensor can obtain $\bsf\left\lceil\frac{\msf}{\bsf}\right\rceil-\bsf$ sensing signals where $k=1,2,...,t_0 -\bsf$, satisfying \([\xv_1^{[0]}, \xv_2^{[0]}, \dots, \xv_{\bsf\left\lceil\frac{\msf}{\bsf-1}\right\rceil -\bsf}^{[0]}] [\xv_1^{[0]}, \xv_2^{[0]}, \dots, \xv_{\bsf\left\lceil\frac{\msf}{\bsf-1}\right\rceil -\bsf}^{[0]}]^H= I\). We can obtain the sensing signal $\xv_s(t)$ to be transmitted by the \(k\)-th antenna at the transmitter as, 

\begin{equation}
    \begin{aligned}
        \begin{bmatrix}
            x_{(1,k)} + x_{(2,k)} + x_{(3,k)} + \dots + x_{(\bsf-1,k)} \\
            x_{(2,k)} + x_{(3,k)} + \dots + x_{(\bsf-1,k)} \\
            x_{(1,k)} + x_{(3,k)} + \dots + x_{(\bsf-1,k)} \\
            \vdots \\
            x_{(1,k)} + x_{(2,k)} + \dots + x_{(\bsf-2,k)}
        \end{bmatrix}
        \\
        =
        \begin{bmatrix}
            1 & 1 & 1 & \dots & 1 \\
            0 & 1 & 1 & \dots & 1 \\
            1 & 0 & 1 & \dots & 1 \\
            \vdots & \vdots & \vdots & \ddots & \vdots \\
            1 & 1 & 1 & \dots & 0
        \end{bmatrix}
        \begin{bmatrix}
            x_{(1,k)} \\
            x_{(2,k)} \\
            x_{(3,k)} \\
            \vdots \\
            x_{(\bsf-1,k)}
        \end{bmatrix}.
    \end{aligned}
\end{equation}

For the sensor:

\begin{equation*}
    \yv_s^{[\Ksf+1]}(t) = \hv^{[\Ksf+1]}(t) \xv(t)+z_s(t),
\end{equation*}
where \(\hv^{[\Ksf+1]}(t) = \begin{bmatrix} H^{[\Ksf+1,1]}(t) & \dots & H^{[\Ksf+1,\msf]}(t) \end{bmatrix}\). 

Thus, Since the sensor experiences a slow fading process, the channel remains constant over several time slots, and the communication signal remains constant as well. The sensor can subtract the received signal at time slot \(t\) from the received signal at the first time slot:
\begin{align*}
    \yv_s^{[\Ksf+1]}(1) - \yv_s^{[\Ksf+1]}(t) &= \hv^{[\Ksf+1]}(1)\xv(1) - \hv^{[\Ksf+1]}(t)\xv(t) \nonumber \\
    &= \hv^{[\Ksf+1]}(t)\xv_{t-1}^{[0]}+\zv_s(1)-\zv_s(t),
\end{align*}
thereby canceling the interference caused by the communication signals and obtaining an effective observation of \(\xv_{t-1}^{[0]}\). Over \(\bsf\) time slots, a total of \(\bsf-1\) effective observations can be made, and the sensing degrees of freedom (SDoF) is \((\bsf-1)/\bsf\).

The system’s total sensing and communication degrees of freedom are \((\text{SDoF}, \text{CDoF}) = \left(\frac{\left\lceil\frac{\msf}{\sum_{k=1}^{\Ksf}\nsf_k}\right\rceil-1}{\left\lceil\frac{\msf}{\sum_{k=1}^{\Ksf}\nsf_k}\right\rceil}, \frac{\msf}{\left\lceil\frac{\msf}{\sum_{k=1}^{\Ksf}\nsf_k}\right\rceil}\right).\)

\section{Proof of Remark~\ref{rem:extension}}
\subsection{Proof of~\eqref{eq:extension KUD}}
\label{pro:UDnco+1To2}
This is the general description of the proposed TIM scheme for the $(\Ksf, U, D)$ neighboring antidotes network with an adding sensor.
For this framework, 
we propose an ISAC scheme based on TIM achieving the tradeoff point  
$(\text{sDoF}, \text{cDoF}) = \left( \frac{U+1}{\Ksf-D+U}, \frac{\Ksf(U+1)}{\Ksf-D+U} \right)$ in (\ref{eq:extension KUD}). Each receiver is disconnected from the previous $U$ and next $D\geq U$ transmitters, and connected to other $\Ksf-U-D>0$ transmitters. Remind that the sensor is a subset of one communication that the network topology could be pressed as 
\begin{align*}
& \mathcal{R}_c^{[k]} = \{ k, <\negmedspace k + D + 1 \negmedspace>_{\Ksf}, \ldots, <\negmedspace k - U - 1\negmedspace>_{\Ksf}\}, \\
& \Rc_s= \{<\negmedspace i + D + 1 \negmedspace>_{\Ksf}, \ldots, <\negmedspace i - U - 1\negmedspace>_{\Ksf}\},
\end{align*}
for each $i, k\in \{1,\ldots,\Ksf\}$, To simplify notation without loss of generality, we set $i=1$, thus we can get $\Rc_s = \{ 2+D \, \ldots,  \Ksf-U \}$.
Let $t_0=(\Ksf-D+U)\left\lceil\frac{(\Ksf-D+U)}{U+1}\right\rceil$, during these $(\Ksf-D+U) \left\lceil\frac{(\Ksf-D+U)}{U+1}\right\rceil$ time slots, the sensor will obtain $(U+1)\left\lceil\frac{(\Ksf-D+U)}{U+1}\right\rceil= N$ effective observations, through the sensing signals $\xv_1^{[0]},\xv_2^{[0]},...,\xv_{N}^{[0]} \in \CC^{(\Ksf-D-U-1)\times1}$, satisfying  \([\xv_1^{[0]}, \xv_2^{[0]},..., \xv_{N}^{[0]}] [\xv_1^{[0]}, \xv_2^{[0]},..., \xv_{N}^{[0]}]^H= I\). In each period of $(\Ksf-D+U)$ time slots, we let each receiver decode $U+1$ communication messages, and let the sensor obtain $U+1$ effective observations. In the next, we will illustrate our proposed scheme for the first $K-D+U$ time slots, where each receiver $k \in \{1,2,...,\Ksf\}$ should recover $U+1$ messages $W_1^{[k]},\ldots,W_{U+1}^{[k]}\in \CC$. 

Denote  the transmitted signals of transmitter $j\in \{1,\ldots,\Ksf\}$ by $\xv^{[j]} =[x^{[j]}(1),\ldots,x^{[j]}(\Ksf-D+U)]^T \in \mathbb{C}^{(\Ksf-D+U) \times 1} $.
For the  transmitters connected to the sensor, we design based on the cyclic coding,
\begin{align*}
    &\xv^{[i]} \negmedspace= \negmedspace\vv^{[i]} W_1^{[i]}\negmedspace + \negmedspace\vv^{[<i+1>_{\Ksf}]} W_2^{[i]}\negmedspace +\negmedspace \ldots\negmedspace +\negmedspace \vv^{[<i+U>_{\Ksf}]} W_{U+1}^{[i]} + \xv_s^{[i]}, 
\end{align*}
for each $i\in \Rc_s $, where  $\xv_s^{[j]}, j\in \mathcal{R}_s$ are the sensing signals to be determined later. For the
  transmitters not connected to the sensor,  design based on the cyclic coding,
\begin{align*}
\xv^{[i]} \negmedspace= \negmedspace\vv^{[i]} W_1^{[i]}\negmedspace + \negmedspace\vv^{[<i+1>_{\Ksf}]} W_2^{[i]}\negmedspace +\negmedspace \ldots\negmedspace +\negmedspace \vv^{[<i+U>_{\Ksf}]>} W_{U+1}^{[i]},
\end{align*}
for each $i\in \{1,\ldots,\Ksf \} \setminus \Rc_s$. Note that 
 $\vv^{[1]},...,\vv^{[\Ksf]}\in \CC^{\Ksf-D+U}$ represent $\Ksf$ random i.i.d vectors, where any $\Ksf-D+U$ of them are linearly independent. 
 
 For each $k\in \{1,\ldots,\Ksf\}$,  by removing the dedicated sensing signals from the received signals, the \(k\)-th  receiver obtains 
\begin{align*}
  &  \hat{\yv}^{[k]} = \negmedspace\negmedspace \sum_{i \in \Rc^{[k]}_c \backslash \{k\} } \negmedspace\negmedspace H^{[ki]}(\vv^{[i]} W_1^{[i]} \negmedspace+\negmedspace \ldots\negmedspace +\negmedspace \vv^{[<i+U>_{\Ksf}]} W_{U+1}^{[i]})\negmedspace +\negmedspace\zv^{[k]} \\
    &= \underbrace{H^{[kk]}(\vv^{[k]} W_1^{[k]}\negmedspace + \negmedspace\vv^{[<k+1>_{\Ksf}]} W_2^{[k]}\negmedspace +\negmedspace \ldots \negmedspace+\negmedspace \vv^{[<k+U>_{\Ksf}]} W_{U+1}^{[k]})}_{\text{desire signals}} \\
    &+ \underbrace{\sum^{k+\Ksf-1}_{i=k+D+1}\vv^{[<i>_{\Ksf}]}\sum^{min(\Ksf -U-D,i)}_{s=max(1,i-U)}H^{[ks]}W_{i-s+1}^{[s]}}_{\text{interference signals}} +\zv^{[k]},
\end{align*} 
where \(\max(a, b)\) is the larger value (i.e., \(a\) if \(a \geq b\), otherwise \(b\)) and \(\min(a, b)\) is the smaller value (i.e., \(a\) if \(a \leq b\), otherwise \(b\)). Therefore, the dimension of the interference signals is aligned to \( \Ksf - D -1 \), by the linear independence of $\Ksf -D -1$ vectors $\vv^{[<k+D+1>_{\Ksf+1}]}\ \vv^{[<k+D+2>_{\Ksf+1}]}\ \ldots \vv^{[<k+\Ksf>_{\Ksf+1}]}$, receiver $k$ can recover \( U+1 \) messages from the first \(\Ksf-D+U\) time slots. Thus, the communication degree of freedom is \(\frac{\Ksf(U+1)}{\Ksf-D+U}\).

For the sensor, the received signal is:
\begin{align*}
  &  \yv^{[\Ksf+1]} = \sum^{\Ksf-U }_{i=D+2 }H^{[\Ksf+1,i]}\xv^{[i]} + \zv^{[\Ksf+1]} \nonumber \\
       &=\sum^{\Ksf-D-1}_{i=1}\vv^{[i]}\sum^{min(\Ksf-U-D,i)}_{s=max(1,i-U)}H^{[\Ksf+1,s]}W_{i-s+1}^{[s]} \\
       &+\sum^{\Ksf-U}_{i=D+2}H^{[\Ksf+1,i]}\xv_s^{[i]} \negmedspace + \negmedspace \zv^{[\Ksf+1]}.
\end{align*}

The interference dimension from communication signals is the same as that for communication users, i.e., \(\Ksf-D-1\). Therefore, we can get $U+1$ left null vectors $\vv_{0,1},\vv_{0,2},\dots,\vv_{0,U+1}$ in $\begin{bmatrix} \vv^{[D+2]} \ldots  \vv^{[\Ksf]}\end{bmatrix}$ to remove the communication interference within \(\Ksf-D+U\) time slots, thus
\begin{align*}
  \vv_{0,u} \yv^{[\Ksf+1]}\
  &= \negmedspace\sum_{i=D+2}^{\Ksf-U} H^{[\Ksf+1,i]} \vv_{0,u} \xv^{[i]} \negmedspace + \negmedspace \vv_{0,u} \zv^{[\Ksf+1]} \\
    &= \sum_{i=D+2}^{\Ksf-U} H^{[\Ksf+1,i]} \vv_{0,u} \xv_s^{[i]} + z_u'^{[\Ksf+1]} .
\end{align*}

Let $x_u^{[0]}=\begin{bmatrix} \vv_{0,u}\xv_s^{[D+2]} & \cdots & \vv_{0,u}\xv_s^{[\Ksf-U]} \end{bmatrix}^H$, in other $(\Ksf-D-U)$ transmission, we can use the same way to get $x_{U+2}^{[0]},x_{U+3}^{[0]},\cdots,x_{\Ksf-D-U}^{[0]}$, satisfying \( [\xv_1^{[0]}, \xv_2^{[0]}, \cdots,x_{\Ksf-D-U}^{[0]}][\xv_1^{[0]}, \xv_2^{[0]}, \cdots,x_{\Ksf-D-U}^{[0]}]^H = I \). In other words, we solve $\xv_s^{[D+2]}$ with $\Ksf-D+U$ parameters satisfying $\vv_{0,u}\xv_s^{[D+2]}=x_{u,1}$, where $x_{u,1}$ is the first element of $\xv_u^{[0]}$, and similarly for $\xv_s^{[D+3]}, \dots, \xv_s^{[\Ksf-U]}, \xv_s^{[\Ksf+1]}$. So one effective observation of the channel can be obtained, achieving the sDoF equal to \(\frac{U+1}{\Ksf-D+U}\).

\subsection{Proof of~\eqref{eq:coo topo2}}
\label{pro:KDco+1}
This is the general description of the proposed TIM scheme for the $(\Ksf, d)$ regular network with an adding sensor. In this framework,
we propose an ISAC scheme based on TIM achieving the tradeoff point  
$(\text{sDoF}, \text{cDoF}) = \left( \frac{1}{d+1}, \frac{2\Ksf}{d+1} \right)$.

Recall that the set of connected transmitters of the sensor is a subset of one communication  that the network topology could be expressed as
\begin{align*}
    & \Rc^{[k]}_c = \{k, <\negmedspace k+1 \negmedspace>_{\Ksf}, \ldots, <\negmedspace k+d-1 \negmedspace>_{\Ksf} \}, \\
    & \Rc_s = \{<\negmedspace i+1 \negmedspace>_{\Ksf}, \ldots, <\negmedspace i+d-1 \negmedspace>_{\Ksf}\},
\end{align*}
for each $i, k\in \{1,\ldots,\Ksf\}$, To simplify notation without loss of generality, we set $i=1$, thus we can get $\Rc_s = \{ 2 \, \ldots,  d \}$.

Let $t_0 = (d+1)(d+1)$. In $(d+1)(d+1)$ time slots, the sensor will obtain $d+1$ effective observations, through the sensing signals $\xv_1^{[0]},\xv_2^{[0]},...,\xv_{d+1}^{[0]} \in \CC^{(d+1)\times1}$, satisfying  \([\xv_1^{[0]}, \xv_2^{[0]},..., \xv_{d+1}^{[0]}] [\xv_1^{[0]}, \xv_2^{[0]},..., \xv_{d+1}^{[0]}]^H= I\). In each period of $(d+1)$ time slots, we let each receiver decode $2$ communication messages, and let the sensor obtain one effective observation. In the next, we   illustrate our proposed scheme for the first $d+1$ time slots, where each receiver $k \in \{1,2,...,\Ksf\}$ should recover $2$ messages $W_1^{[k]},W_{2}^{[k]}\in \CC$.

Denote  the transmitted signals of transmitter $j\in \{1,\ldots,\Ksf\}$ by $\xv^{[j]} =[x^{[j]}(1),\ldots,x^{[j]}(d+1)]^T \in \mathbb{C}^{(d+1) \times 1} $.
For the $d$ transmitters  connected to the sensor, let
\begin{align*}
    & \xv^{\left[j\right]} \negmedspace=\negmedspace \vv^{\left[j\right]} W_1^{\left[j\right]} \negmedspace+\negmedspace \vv^{\left[j+1\right]} W_2^{\left[j-d+1\right]} \negmedspace+\negmedspace \xv_s^{\left[j\right]}, \forall j\in \{2,\ldots,d\}.
\end{align*}
For the transmitters in the set $\{1, d+1, \ldots, \Ksf\}$, which are not connected to the sensor, let
\begin{align*}
    &\xv^{\left[j\right]} = \vv^{\left[j\right]} W_1^{\left[j\right]} + \vv^{\left[j+1\right]} W_2^{\left[j-d+1\right]}, \forall j\in \{1, d+1, \ldots, \Ksf\}. 
\end{align*}

By removing the dedicate sensing signal, what the $k\in \{1,\ldots,\Ksf\}$-th  communication receiver received is:

\begin{align*}
    \hat{\yv}^{\left[k\right]} &= \sum_{j=k}^{k+d-1} H^{\left[kj\right]} \left( \vv^{\left[j\right]} W_1^{\left[j\right]} + \vv^{\left[j+1\right]} W_2^{\left[j-d+1\right]} \right)  + \zv^{[k]} \\
&=\underbrace{H^{\left[kk\right]} \vv^{\left[k\right]} W_1^{\left[k\right]} + H^{\left[k,k+d-1\right]} \vv^{\left[k+d\right]} W_2^{\left[k\right]}}_{{\rm desired \  signal}} \\
    \quad &+ \sum_{j=k+1}^{k+d-1} \vv^{\left[j\right]} \left( H^{\left[kj\right]} W_1^{\left[j\right]} + H^{\left[k,j-1\right]} W_2^{\left[j-d\right]} \right)+ \zv^{[k]}.
\end{align*}
The dimension of the desired signals  is 2, and the dimension of the interference signals  is $d-1$, we can achieve symmetric $\text{cDoF} = \frac{2}{d+1}$ for each receiver. 
Thus, the cDoF is $\frac{2\Ksf}{d+1}$.

For the sensor, the received signal $\yv^{\left[s\right]}$ is:
\begin{align*}
    \yv^{[s]} &= \sum_{j=2}^{d} H^{[\Ksf+1,j]} \left( \vv^{[j]} W_1^{[j]} \negmedspace+ \negmedspace\vv^{[j+1]} W_2^{[j-d+1]} \negmedspace+ \negmedspace\xv_s^{[j]} \right) \negmedspace+\negmedspace \zv^{[\Ksf+1]} \\
    &=\sum^{d}_{j=2}\vv^{[j]}(H^{[\Ksf+1,j]}W_1^{[j]}+H^{[\Ksf+1,j-1]}W_2^{[j-d]})+\zv^{[\Ksf+1]}.
\end{align*}

Note that $[\vv^{\left[2\right]}  \vv^{\left[3\right]}  \cdots  \vv^{\left[d+1\right]}]$ with dimension $(d+1)\times(d)$ has one linearly independent left null vectors with high probability; let $\vv_0 \ in \CC^{1 \times d+1}$ be the non-zero left null vector. Thus 
\begin{align*}
    v_0 \yv^{\left[s\right]} \negmedspace=\negmedspace \sum_{j=2}^{d}\negmedspace\negmedspace H^{\left[\Ksf+1,j\right]}  v_0 \xv_s^{\left[j\right]}  \negmedspace+\negmedspace  H^{\left[\Ksf+1,\Ksf+1\right]} v_0 \xv_s^{\left[\Ksf+1\right]} \negmedspace+\negmedspace\vv_0\zv^{[\Ksf+1]} .
\end{align*}

Let $x_1^{\left[0\right]} = \left[ v_0 \xv_s^{\left[2\right]} , v_0 \xv_s^{\left[3\right]} , \cdots , v_0 \xv_s^{\left[d\right]} , v_0 \xv_s^{\left[\Ksf+1\right]} \right]$, and this gives one valid observation.  So one effective observation of the channel can be obtained, achieving a sDoF of \(\frac{1}{d+1}\).

\bibliographystyle{IEEEtran}
\bibliography{reference}

\end{document}